\documentclass[twocolumn,prb,showpacs,floatfix]{revtex4}
\usepackage{graphicx}
\usepackage{bm}

\begin{document}

\title{\textbf{Observation of extended scattering continua
characteristic of spin fractionalization in the 2D frustrated
quantum magnet Cs$_2$CuCl$_4$ by neutron scattering}}
\author{R. Coldea$^{1,2,3}$, D.A. Tennant$^{1,3}$, and Z. Tylczynski$^{4}$}
\affiliation{$^1$Oxford Physics, Clarendon Laboratory, Parks Road,
Oxford OX1 3PU, United Kingdom\\
$^2$Solid State Division, Oak Ridge National Laboratory, Oak
Ridge, Tennessee 37831-6393\\
$^3$ISIS Facility, Rutherford Appleton Laboratory, Chilton, Didcot
OX11 0QX, United Kingdom\\
$^4$Institute of Physics, Adam Mickiewicz University, Umultowska
85, 61-614 Poznan, Poland}

\date{\today}


\begin{abstract}
The magnetic excitations of the quasi-2D spin-1/2 frustrated
Heisenberg antiferromagnet Cs$_2$CuCl$_4$ are explored throughout
the 2D Brillouin zone using high-resolution time-of-flight
inelastic neutron scattering. Measurements are made both in the
magnetically-ordered phase, stabilized at low temperatures by the
weak inter-layer couplings, as well as in the spin liquid
phase above the ordering temperature $T_N$, when the 2D magnetic
layers are decoupled. In the spin liquid phase the dynamical
correlations are dominated by highly-dispersive excitation
continua, a characteristic signature of fractionalization of $S$=1
spin waves into pairs of deconfined $S$=1/2 spinons and the hallmark
of a resonating-valence-bond (RVB) state. The boundaries of the
excitation continua have strong 2D-modulated incommensurate dispersion
relations. 
Upon cooling below $T_N$ magnetic order in an incommensurate
spiral forms due to the 2D frustrated couplings. In this phase
sharp magnons carrying a small part of the total scattering weight
are observed at low energies, but the dominant continuum
scattering which occurs at medium to high energies is essentially
unchanged compared to the spin liquid phase. Linear spin-wave
theory including one- and two-magnon processes can describe the
sharp magnon excitation, but not the dominant continuum
scattering, which instead is well described by a parameterized
two-spinon cross-section. 
Those results suggest a cross-over in the nature of the
excitations from $S$=1 spin waves 
at low energies to deconfined $S$=1/2 spinons at medium to high
energies, which could be understood if Cs$_2$CuCl$_4$ was in the
close proximity of transition between a fractional RVB spin liquid
and a magnetically ordered state. A large renormalization factor
of the excitation energies ($R$=1.63(5)), indicating strong
quantum fluctuations in the ground state, is obtained using the
exchange couplings determined from saturation-field measurements.
We provide an independent consistency check of this quantum
renormalization factor using measurements of the second moment of
the paramagnetic scattering.
\end{abstract}

\maketitle

\section{Introduction}
\label{sec_introduction}

One of the most remarkable phenomena that can occur in strongly
correlated systems is the emergence of particles with fractional
quantum numbers. This occurs when fluctuations in the many-body
quantum ground state are strong enough to provide a screening by
which fractional components {\em de-confine} from local integer
constraints. The best known example in magnetism is the
one-dimensional(1D) $S$=1/2 Heisenberg antiferromagnetic chain
(HAFC) where a semi-classical $S$=1 spin wave spontaneously decays
into a pair of deconfined $S$=1/2 spinons that separate away
independently\cite{Muller81,Haldane93,Tennant95a}. The signature
of spin fractionalization is a highly-dispersive continuum of
excited states instead of sharp single-particle poles in the
dynamical correlations, probed directly by inelastic neutron
scattering experiments.

Fractionalization arises from subtle many-body quantum correlation
effects (in the spin-1/2 HAFC chain a topological Berry phase term
is responsible\cite{Haldane85}) and higher dimensional
realizations of such exotic physics have been eagerly sought.
Amongst the proposed theoretical scenarios is the
Resonating-Valence-Bond (RVB) spin liquid state introduced by
Anderson\cite{Anderson73} in the context of 2D frustrated quantum
magnets. Here spins spontaneously pair into singlet bonds which
fluctuate between many different configurations; breaking a
singlet bond releases two $S$=1/2 spinons that propagate away
independently in the background of the fluctuating bonds. Central
to the stability of such an RVB phase is frustration, which
enhances quantum fluctuations over the mean-field effects that
would otherwise favor conventional ordered phases.

Nearly all experimentally studied phases in 2D quantum magnets
have been found to show conventional confined phases, which are
well characterized. For example, in {\em un}frustrated magnets
typified by the $S$=1/2 Heisenberg antiferromagnet on a square
lattice (HSL) mean-field effects dominate and quantum fluctuations
cause only small renormalizations to the semi-classical
description of a N\'{e}el ordered ground state with $S$=1
transverse spin-wave excitations \cite{Hayden91}. Other 2D quantum
phases include dimerized states recently observed experimentally
\cite{Kageyama99} where spins are paired into singlet bonds
(dimers) arranged in a regular (fixed) ordered pattern in the
ground state; such a phase has a spin gap to a triplet of $S$=1
magnon excitations, i.e. spinons are confined.

However, spins in frustrated geometries are believed to behave in
much more unconventional ways, and interest in such systems is
also motivated by the observation of superconductivity (possibly
mediated by spin fluctuations) in charge-doped triangular lattice
materials $\kappa$-(BEDT-TTF)$_2$X \cite{McKenzie98} and
Na$_x$CoO$_2$$\cdot y$H$_2$0 \cite{Takada03}. In this respect the
recent discovery of a 2D quantum magnet (Cs$_2$CuCl$_4$) with
spins on a triangular lattice which shows fractionalization is of
clear importance for elucidating the
underlying magnetism\cite{Coldea01}.   
Indeed, deconfined phases in 2D and the conditions required to
create them are a major unsolved theoretical
problem\cite{Anderson73,Kalmeyer87,Read91,Nersesyan02,Chung01,Zhou02,Senthil00}.
Cs$_2$CuCl$_4$ is a quasi-2D $S$=1/2 frustrated Heisenberg
antiferromagnet on an anisotropic triangular
lattice\cite{Coldea02} and neutron scattering measurements show
the dynamical correlations to be dominated by a broad continuum of
excited states\cite{Coldea01} as characteristic of deconfined
$S$=1/2 spinons. Here we build on our initial neutron scattering
results and provide comprehensive measurements throughout the
Brillouin zone. Effects of the strong two-dimensionality are
explicitly observed at all energy scales of the excitations.
Measurements are made both at temperatures above $T_N$ when the 2D
magnetic layers are decoupled as well as below $T_N$ when
mean-field effects from the weak inter-layer couplings stabilize
3D magnetic order with an incommensurate spiral structure.

The paper is organized as follows. The crystal structure and
magnetism of Cs$_2$CuCl$_4$ are described in Sec.\
\ref{intro_cscucl} and the experimental technique used to probe
the excitations is explained in Sec.\ \ref{sec_experiment}. The
dispersion relation and scattering lineshapes measured in the
low-temperature ordered phase are presented next in Secs.\
\ref{subsec_orderedphase_dispersions} and
\ref{subsec_orderedphase_lineshapes}. We find sharp magnon peaks
carrying a small part of the total scattering weight at low
energies and highly-dispersive continua carrying the majority of
the scattering weight at moderate to high energies. Results are
first compared to linear spin-wave theory (reviewed in Appendix A)
including both one and two-magnon processes in Sec.\
\ref{subsubsec_lineshape_lswt}. This theory is found inadequate to
describe the dominant continuum scattering, which instead is well
described by a parameterized two-spinon cross-section (Sec.\
\ref{subsubsec_lineshape_2spinon}). Measurements in the spin
liquid phase above $T_N$ where the magnetic layers are decoupled
are shown in Sec.\ \ref{subsec_spinliquid}. The paramagnetic
scattering is described in Sec\ \ref{subsec_paramagnetic} where
the extracted second moment is compared with sum rules. In Sec.\
\ref{subsec_disc_crossover} we discuss issues in depth with
reference to proximity to a spinon confinement transition. The
paramagnetic scattering is discussed in the context of spinon
systems at high temperatures in Sec.\ \ref{subsec_disc_paramag}.
Finally, the main results and conclusions are summarized in Sec\
\ref{sec_conclusions}.

\section{Crystal structure and magnetic properties of Cs$_2$CuCl$_4$}
\label{intro_cscucl}

\begin{figure}[t]
\begin{center}
  \includegraphics[width=6cm,bbllx=165,bblly=58,bburx=518,
  bbury=826,angle=0,clip=]{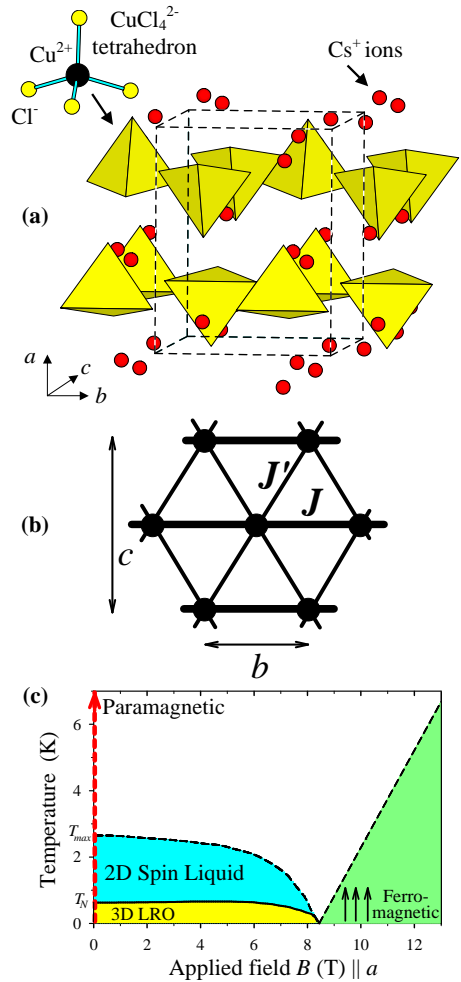}
  \caption{\label{fig_cs2cucl4_structure} (color online)
  (a) Crystal structure in Cs$_2$CuCl$_4$ showing the
  CuCl$_{4}^{2-}$ tetrahedra (pyramids) arranged in
  layers ($bc$ plane). The orthorhombic unit cell is
  indicated by the dashed rectangular box.
  (b) Magnetic exchange paths in a ($bc$) layer form a
  two-dimensional anisotropic triangular lattice: strong
  bonds $J$ (heavy lines $\parallel b$) and smaller
  frustrating zig-zag bonds $J^{\prime}$ (thin lines).
  (c) Schematic phase diagram of Cs$_2$CuCl$_4$ in
  temperature and magnetic field along $a$ showing
  the region probed by the present experiments (dashed
  vertical arrow at $B=0$). The magnetic phases are:
  3D LRO ($T<T_N$) with spiral magnetic long-range order,
  spin-liquid ($T_N<T<T_{max}$) characterized by strong
  intra-layer antiferromagnetic correlations (at
  $T_{max}$=2.65 K the magnetic susceptibility has a maximum),
  paramagnetic ($T \gg (J,J^{\prime})$) and ferromagnetic ($B>B_C$)
  where spins are ferromagnetically aligned by the applied
  field. The solid line is a phase transition boundary
  and dashed lines show cross-overs.}
\end{center}
\end{figure}

The crystal structure of Cs$_2$CuCl$_4$ is
orthorhombic\cite{Bailleul91}({\em Pnma}) with lattice parameters
$a$=9.65 \AA\ , $b$=7.48 \AA\ and $c$=12.26 \AA\ at 0.3 K. The
structure is illustrated in Fig.\ \ref{fig_cs2cucl4_structure}(a)
and consists of CuCl$_4$$^{2-}$ tetrahedra arranged in layers
($bc$ plane) separated along $a$ by Cs$^{+}$ ions. The material is
an insulator with each Cu$^{2+}$ ion carrying a spin of 1/2.
Crystal field effects quench the orbital angular momentum
resulting in near-isotropic Heisenberg spins on each Cu$^{2+}$
ion. There are four such ions in each orthorhombic unit cell, two
located on each CuCl layer as illustrated in Fig.\
\ref{fig_cs2cucl4_structure}(a).

Our previous measurements\cite{Coldea02} showed that Cs$_2$CuCl$_4$
is a quasi-2D low-exchange quantum magnet. This quasi-2D character
is a result of the layered crystal structure, which restricts the
main superexchange routes to neighboring spin sites in the ($bc$)
plane. Since this superexchange route is mediated by two
nonmagnetic Cl$^-$ ions the exchange energies are low $\sim$1-4 K.
In each layer the exchange paths form a triangular lattice with
non-equivalent couplings indicated in Fig.\
\ref{fig_cs2cucl4_structure}(b): exchange $J$ along the $b$-axis
and $J^{\prime}$ along the zig-zag bonds. The main 2D Hamiltonian
describing Cs$_2$CuCl$_4$ is thus
\begin{equation}
{\cal H}=\sum\limits_{\langle i,i^{\prime }\rangle }J{\bm
S}_{i}\cdot {\bm S}_{i^{\prime }}+\sum\limits_{\langle i,j\rangle
}J^{\prime }{\bm S}_{i}\cdot {\bm S}_{j},  \label{eq_ham}
\end{equation}
with each interacting spin-pair counted once. This Hamiltonian
interpolates between non-interacting chains ($J^{\prime }=0$), the
fully frustrated triangular lattice ($J^{\prime }=J$), and
unfrustrated square lattice ($J=0$). Since the main spin couplings
define a 2D isosceles triangular lattice as shown in Fig.\
\ref{fig_cs2cucl4_structure}(b), measurements of the excitations
will be discussed in terms of the 2D Brillouin zone of this
triangular lattice, even though the full crystal symmetry is
orthorhombic.

\subsection{Hamiltonian}
\label{intro_cscucl_hamiltonian}
The full spin Hamiltonian of Cs$_2$CuCl$_4$ and its parameters
have been previously determined\cite{Coldea02} from measurements
in high applied magnetic fields that overcome the
antiferromagnetic couplings and stabilize the fully-aligned
ferromagnetic ground state $| \! \uparrow \uparrow \uparrow \!
\ldots \rangle$ in Fig.\ \ref{fig_cs2cucl4_structure}(c). In this
unique phase quantum fluctuations are entirely suppressed by the
large field, the excitations are magnons and their dispersion
relations image directly the Fourier transform of the bare (i.e.
unrenormalized) exchange couplings. Using this method the
determined interactions are: $J$=0.374(5) meV along the $b$-axis
and $J^{\prime}/J$=0.34(3) along the zig-zag bonds, the couplings
in Fig.\ \ref{fig_cs2cucl4_structure}(b), and small
Dzyaloshinskii-Moriya terms along the zig-zag bonds
$D_a/J$=0.053(5) and inter-layer couplings
$J^{\prime\prime}/J$=0.045(5) (for further details see Ref.\
\onlinecite{Coldea02}). Note that sum rules to be discussed in
Sec.\ \ref{subsec_paramagnetic} also provide an independent
verification of the strength of the couplings.

The finite interlayer couplings ($J^{\prime\prime}$) stabilize
long-range magnetic order\cite{Coldea96} below $T_N$=0.62(1) K.
The order is incommensurate due to the frustrated couplings in the
($bc$) plane with the ordering wavevector $Q=(0.5
+\epsilon_0)\bm{b}^*$, $\epsilon_0$=0.030(2). Ordered spins rotate
in a spiral nearly contained in the ($bc$) plane due to the small
DM couplings, which create an effective easy-plane anisotropy. The
projection of the spiral order onto the ($bc$) plane is
schematically illustrated in Fig.\ \ref{fig_spin_rotation}(b).

\section{Experimental details}
\label{sec_experiment}

The samples used were high-quality, large single crystals of
Cs$_2$CuCl$_4$ grown from solution\cite{Soboleva81} and cut into
plates of typical size $2\times2\times0.6$ cm$^3$ optimized for
neutron absorption. Temperature control was provided by a dilution
refrigerator insert with a 0.04 K base temperature. Measurements
were made in the different regions of the magnetic phase diagram
indicated in Fig.\ \ref{fig_cs2cucl4_structure}(c) (zero-field
line): the 3D spiral ordered phase (below 0.1 K), the spin liquid
phase above $T_N$ (at 0.75 and 0.9 K) and also the paramagnetic
phase (at 12.8 and 15 K).

\begin{figure}[t]
\begin{center}
  \includegraphics[width=7.3cm,bbllx=77,bblly=103,bburx=472,
  bbury=738,angle=0,clip=]{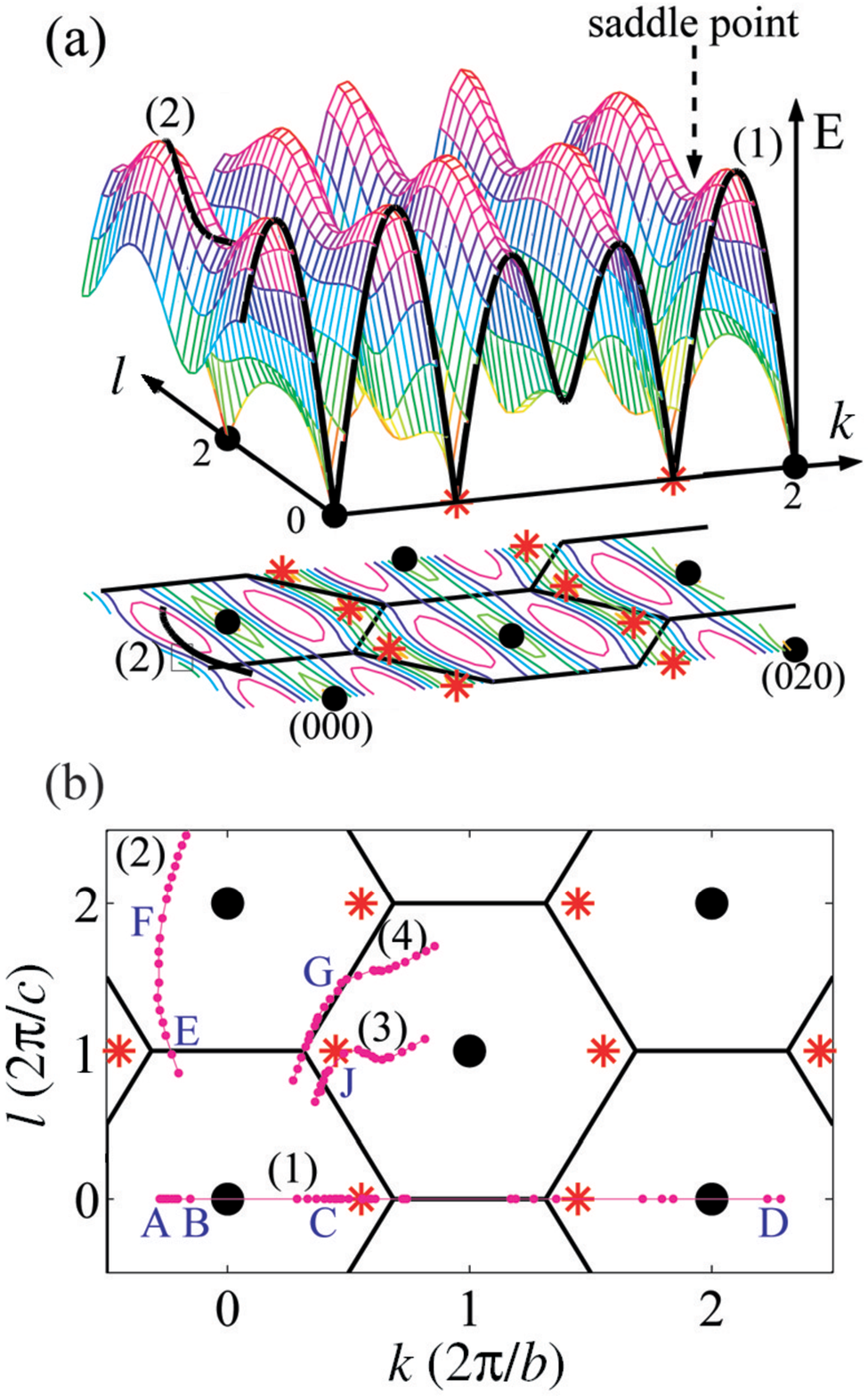}
  \caption{\label{fig_dispersion_surface} (color online)
  (a) Surface plot of
  the dispersion relation in Eq.\ (\ref{eq_dispersion}) as
  a function $k$ and $l$ with constant-energy contours shown
  projected onto the basal plane (thin curves). (b) Reciprocal
  space diagram showing where measurements were made. Numbers
  (1) to (4) refer to directions (curves) in reciprocal space
  where dispersion points (solid dots) were extracted (curves
  (1) and (2) are also indicated by bold curves on the surface
  plot in (a)); the extracted dispersions are plotted in the
  numbered panels in Fig.\ \ref{fig_dispersion_alongkl}.
  Capital letters A-J indicate location of energy scans.
  Thick lines show the near-
  hexagonal Brillouin zones of the 2D triangular lattice
  in Fig.\ \ref{fig_cs2cucl4_structure}(b), filled circles
  are zone centers ($\bm \tau$) and stars at
  $\bm \tau \pm \bm Q$, ${\bm Q}=(0.5+\epsilon_0){\bm b}^*$
  mark incommensurate magnetic Bragg peaks from the spiral
  order at $T<T_N$.}
\end{center}
\end{figure}

The magnetic excitations were measured using the indirect-geometry
time-of-flight spectrometer IRIS\cite{Carlile92} at the ISIS
spallation neutron source in the UK, which combines high energy
resolution with a large detector solid angle that allows
simultaneous coverage of extended (wavevector, energy) regions.
The detector array collects data on a near-cylindrical surface in
$(k_x,k_y,E)$ space $\left(k_x-\sqrt{(E_f+E)/\alpha}\right)^2 +
k_y^2 = E_f/\alpha$, where $k_x$ (along the incident beam
direction) and $k_y$ are the components of the wavevector transfer
in the horizontal scattering plane, $E$ is the energy transfer,
$E_f$ is the fixed final energy and $\alpha=5.072$ meV \AA$^2$.
Dispersion points are obtained when this measurement surface
intersects the dispersion surface of the magnetic excitations (a
typical intersection is shown by the bold curve (2) in Fig.\
\ref{fig_dispersion_surface}(a)). By changing the orientation of
the crystal the measurement surface can be positioned to probe the
excitations near particular directions in reciprocal space.
Several non-equivalent directions (curves) in the Brillouin zone
were probed and are indicated by the dotted lines numbered (1) to
(4) in Fig.\ \ref{fig_dispersion_surface}(b).  Typical counting
times for one crystal orientation and temperature ranged between
30 to 54 hours at an average ISIS proton current of 170 $\mu$A.
Capital letters A-K indicate location of energy scans with
parameters listed in Table.\ \ref{table_scans}. Throughout this
paper the wavevector transfer $\bm{k}=(h,k,l)$ is expressed in
reciprocal lattice units (rlu)\cite{sl} of ($2\pi/a$, $2\pi/b$,
$2\pi/c$).

\begin{table*}
\begin{center}
\caption{\label{table_scans}Scan parameters: wavevector
$\bm{k}=(h,k,l)$ and energy ($E$) change along scan direction,
location of main peak $\omega_{\bm{k}}$ and polarization factors
($p_x$,$p_z$) at this point (see Eq.\
(\ref{eq_app_polarizations})).}
\begin{ruledtabular}
\begin{tabular}{cccccccc}
Scan&$h$(rlu)&$k$(rlu)&$l$(rlu)&$E$(meV)&$\omega_{\bm{k}}$(meV)&$p_x$&$p_z$\\
\hline
A & $1.31+0.29E-0.025E^2$ & $-0.389 +0.189E-0.016E^2$ & 0                     & $E$ & 0.69(2)  & 1.95 & 0.05 \\
B & $1.69+0.29E-0.025E^2$ & $-0.30  +0.189E-0.015E^2$ & 0                     & $E$ & 0.62(2)  & 1.98 & 0.02 \\
C & $0.84$                & $ 0.21  +0.297E -0.026E^2$  & 0                     & $E$ & 0.56(2)  & 1.75 & 0.25 \\
D & $0.69$                & $2.11   +0.29E -0.025E^2$ & 0                     & $E$ & 0.68(2)  & 1.05 & 0.95 \\
E & 0                     & $-0.33  +0.19E-0.015E^2$  & $0.78+0.37E-0.03E^2$  & $E$ & 0.56(1)  & 1    & 1 \\
F & 0                     & $-0.39  +0.19E-0.02E^2$   & $1.66+0.37E-0.035E^2$ & $E$ & 0.67(2)  & 1    & 1 \\
G & 0                     & 0.5                       & $1.53-0.32E-0.1E^2$  & $E$ & 0.107(10)& 1    & 1 \\
H & 0                     & $0.28   +0.29E -0.025E^2$ & 1.205                 & $E$ & 0.34(2)  & 1    & 1 \\
I & 0                     & $k$                       & $-0.23+3.63k-1.64k^2$ & 0.85(5) & $-$  & 1    & 1 \\
J & 0                     & 0.47                      & $1.0 -0.45 E$         & $E$ & 0.10(1)  & 1    & 1 \\
K & 0                     & $0.29  +0.29E -0.03E^2$   & $0.77-0.14E+0.013E^2$ & $E$ & 0.32(2)  & 1    & 1 \\
\end{tabular}
\end{ruledtabular}
\end{center}
\end{table*}

An energy resolution $\Delta E$=0.019(1) meV
(full-width-at-half-maximum) \cite{abplane} on the elastic line
was achieved using a fixed final energy $E_f$=1.846 meV selected
by pyrolytic graphite (002) analyzers, cooled to liquid Helium
temperatures to reduce the thermal diffuse background. The
detector array had 51-elements covering the range of scattering
angles $25.75^{\circ} \le 2\theta \le 158.0^{\circ}$. The raw
neutron counts vs. time-of-flight were converted into scattering
cross-section intensities $S(\bm{k},\omega)$ [typical data is
shown in Fig.\ \ref{fig_lineshaped}], which were further corrected
for neutron absorption effects using a numerical calculation for a
plate-shaped crystal calibrated against the measured incoherent
quasi-elastic scattering from the sample. The non-magnetic
background was estimated from measured intensities near places in
the Brillouin zone such as the ferromagnetic zone center where the
magnetic scattering contribution is negligible; the accuracy of
this background subtraction procedure is illustrated in Fig.\
\ref{fig_irs_scans}G (solid points).

\section{Results and Analysis}
\label{sec_results}
\subsection{Dispersion relations in the spiral ordered phase below $T_N$}
\label{subsec_orderedphase_dispersions}
\begin{figure*}
\begin{center}
 \includegraphics[width=15cm,bbllx=26,bblly=148,bburx=534,
  bbury=623,angle=0,clip=]{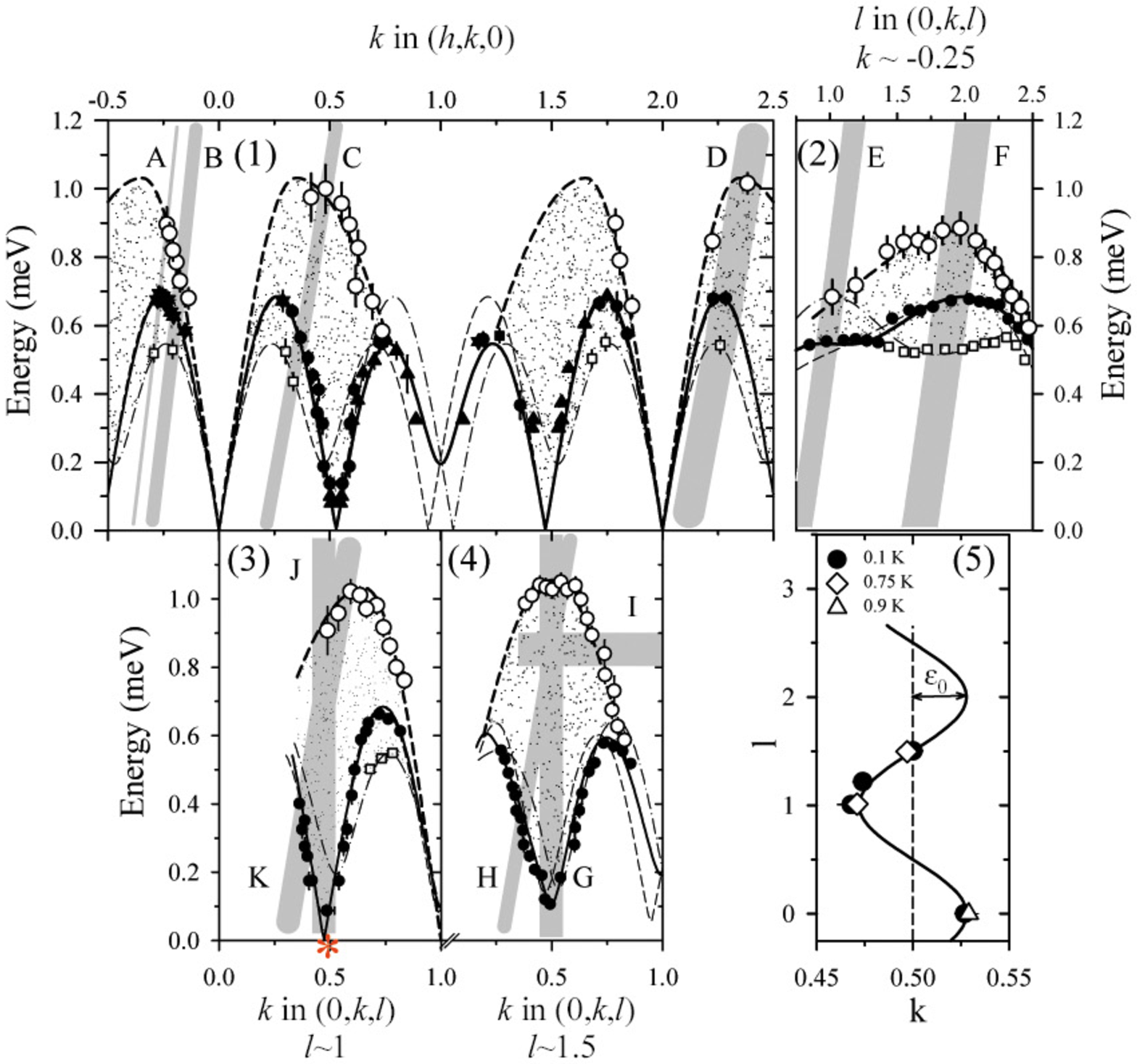}
  {\caption {\label{fig_dispersion_alongkl} Dispersion relation
  of the magnetic excitations at $T<$0.1 K. Numbered panels (1)-(4)
  correspond to directions in the 2D plane indicated in Fig.\
  \ref{fig_dispersion_surface}(b) [(1) is from Ref.\
  \onlinecite{Coldea01}]. Filled symbols are the main peak in
  the lineshape and the solid line is a fit to the spin-wave
  dispersion relation in Eq.\ (\ref{eq_dispersion}); the dotted
  area indicates the extent of the magnetic scattering, large
  open circles mark the experimentally-estimated continuum
  upper boundary (the upper thick dashed line is a guide to the
  eye) and open squares show the lower boundary of the scattering
  (where it could be resolved from the main peak as shown in
  Fig.\ \ref{fig_lineshaped}); dashed and dash-dotted lines are
  the secondary spin-wave dispersions $\omega^+_{\bm{k}}$ and
  $\omega^-_{\bm{k}}$, respectively; light shaded regions
  labelled with capital letters A-K (the line thickness
  represents the wavevector averaging) indicate scan directions.
  Panel (5) shows the incommensurate modulation in the dispersion
  relation vs. $k$ and $l$ compared with predictions
  of Eq.\ (\ref{eq_dispersion}) (solid line) and uncoupled chains
  ($J^{\prime}=0$ dashed line).}}
\end{center}
\end{figure*}

Measurements of the magnetic excitations at the lowest temperature
($T<0.1$ K) in the spiral ordered phase showed strong dispersion
along both directions in the 2D triangular layers ([0$k0]$ and [00$l$]) and
negligible dispersion along the inter-layer direction [$h$00]. Table\
\ref{table_disp} lists key parameters and
Fig.\ \ref{fig_dispersion_surface} shows a surface plot of the
dispersion relation in the 2D triangular plane. Data obtained
from cuts through this dispersion surface along several
non-equivalent directions in the 2D plane [curves numbered (1) to
(4)] is plotted in Fig.\ \ref{fig_dispersion_alongkl}.

\begin{table}
\begin{center}
\caption{\label{table_disp}Dispersion parameters.}
\begin{ruledtabular}
\begin{tabular}{lc}
Zone-boundary energy (meV) along [0$k$0] & \begin{tabular}{cc}$0.67(1)$ and $0.55(1)$ \\ alternating \end{tabular} \\
\begin{tabular}{cc} Zone-boundary energy (meV) along [00$l$] \\ measured at ${\bm k}=(0,0.5,1.5)$ \end{tabular} & $0.107(10)$ \\
Dispersion (meV) along $l$ for $k\sim-0.25$ & $0.12(1)$ \\
Dispersion (meV) along $h$ for $k\sim-0.25$ & $<0.02$ \\
Incommensuration $\epsilon_0$(rlu) & $0.030(2)$  \\
\end{tabular}
\end{ruledtabular}
\end{center}
\end{table}

Fig.\ \ref{fig_dispersion_alongkl}(1) gives the dispersion along
$k$ for $l$=0. Here data points also have a finite component along
$h$, but no dispersion could be detected along this direction for
near-constant $k$, as expected for weak couplings
$J^{\prime\prime}$ between layers. The dispersion along $k$ is
incommensurate, with the minimum energy at the magnetic Bragg peak
positions indexed by the ordering wavevector ${\bm
Q}=(0.5+\epsilon_0){\bm b}^*$, $\epsilon_0$=0.030(2), and has an
asymmetric shape with non-equivalent zone boundary energies of
0.67(1) and 0.55(1) meV on the two sides of the minimum. The
incommensuration and asymmetry of the dispersion are direct
consequences of the frustrated couplings (a symmetric dispersion,
commensurate at the antiferromagnetic point $k$=0.5 would be
expected in the absence of frustration $J^{\prime}$=0 in Fig.\
\ref{fig_cs2cucl4_structure}(b)). The effects of the 2D couplings
are most clearly seen in the dispersion along $l$  for
near-constant $k\sim-0.25$ plotted in Fig.\
\ref{fig_dispersion_alongkl}(2). The observed dispersion is
0.12(1) meV, practically equal to the strength of the couplings
along $l$, $J^{\prime}$=0.128(5) meV as directly measured in Ref.\
\onlinecite{Coldea02}. The significant dispersion observed along
both [0$k$0] and [00$l$] directions indicates a strong 2D
character for the magnetic excitations.

Further measurements that illustrate the $l$-dependence of the
dispersion are shown in Fig.\ \ref{fig_dispersion_alongkl}(3),
with data near $l=1$. Compared to $l$=0 data in (1) the left-right
asymmetry of the dispersion along $k$ has now been reversed and
the minimum energy displaced to the new Bragg peak position in
this zone $k=0.5-\epsilon_0$, as expected by periodicity for a
triangular lattice with a 2D Brillouin zone as shown in Fig.\
\ref{fig_dispersion_surface}. Data near the zone-boundary point
along $l$, $l$=1.5, are plotted in Fig.\
\ref{fig_dispersion_alongkl}(4). Here the dispersion along $k$
appears nearly symmetric around $k$=0.5 with a gap at
$\omega_{(0,0.5,1.5)}$=0.107(10) meV.

The observed modulations in the dispersion relation as a function
of $k$ and $l$ can be well described by the principal spin-wave
dispersion of the 2D frustrated Hamiltonian in Eq.\ (\ref{eq_ham})
(for details see Appendix A)
\begin{equation}
\omega_{\bm k}=\sqrt{\left( J_{\bm k}-J_{\bm Q}\right) \left[
\left( J_{{\bm k}-{\bm Q}}+J_{{\bm k}+{\bm Q}}\right) /2-J_{{\bm
Q}}\right] } \label{eq_dispersion}
\end{equation}
where the Fourier transform of the exchange couplings is $J_{{\bm
k}}=\tilde{J} \cos {2\pi k}+2\tilde{J^{\prime }}\cos {\pi k} \cos
{\pi l}$, ${\bm k} = (h , k ,l)$\cite{sl} and the ordering
wavevector ${\bm Q}=(0.5+\epsilon_0){\bm b}^*$ as above. Solid
lines in Fig.\ \ref{fig_dispersion_alongkl} show the intersection
between the scan directions and Eq.\ (\ref{eq_dispersion}) (using
the actual $k$, $l$ and $E$ at each point in the scan). The {\em
effective} exchange parameters obtained from the above global fit
are $\tilde{J}$=0.61(1) meV and $\tilde{J}^{\prime}$=0.107(10)
meV, in agreement with earlier estimates\cite{Coldea01}
($J$=0.374(5) meV and $J^{\prime}$=0.128(5) meV are the {\em bare}
exchange energies measured using the high-field
technique\cite{Coldea02} where quantum fluctuations are quenched
out). The renormalization of the excitation energy compared to the
classical spin-wave result $\tilde{J} /J$=1.63(5), is very large
and is similar to the exact result $\pi /2$ for the 1D $S$=1/2
HAFC (see e.g. Ref.\ \onlinecite{Muller81}), where it originates
from strong spin-singlet correlations in the ground state. In
contrast, the spin-wave velocity (energy) renormalization in the
unfrustrated $S$=1/2 HSL is only 1.18, see Ref.
\onlinecite{Singh89}. Such a large renormalization indicates
strong quantum fluctuations in the ground state of Cs$_2$CuCl$_4$.

The $l$-dependence of the low-energy dispersion is illustrated in
Fig.\ \ref{fig_dispersion_alongkl}(5), which plots the oscillation
in the minimum energy position vs. $k$ and $l$. Data points were
extracted form fits to the low- to medium-energy part (0.1$<E<$0.4
meV) of the dispersion along $k$ for various $l$'s. The observed
incommensurate oscillation is well explained by Eq.\
(\ref{eq_dispersion}), for which at a given $l$ the minimum energy
occurs at $k=0.5+\frac{1}{\pi}\sin^{-1}\left(
\frac{\tilde{J^{\prime}}}{2\tilde{J}}\cos \pi l \right)$ (solid
line).

\subsection{Excitation lineshapes in the spiral ordered phase below $T_N$}
\label{subsec_orderedphase_lineshapes}
\begin{figure}[t]
\begin{center}
 \includegraphics[width=8cm,bbllx=41,bblly=245,bburx=410,
  bbury=665,angle=0,clip=]{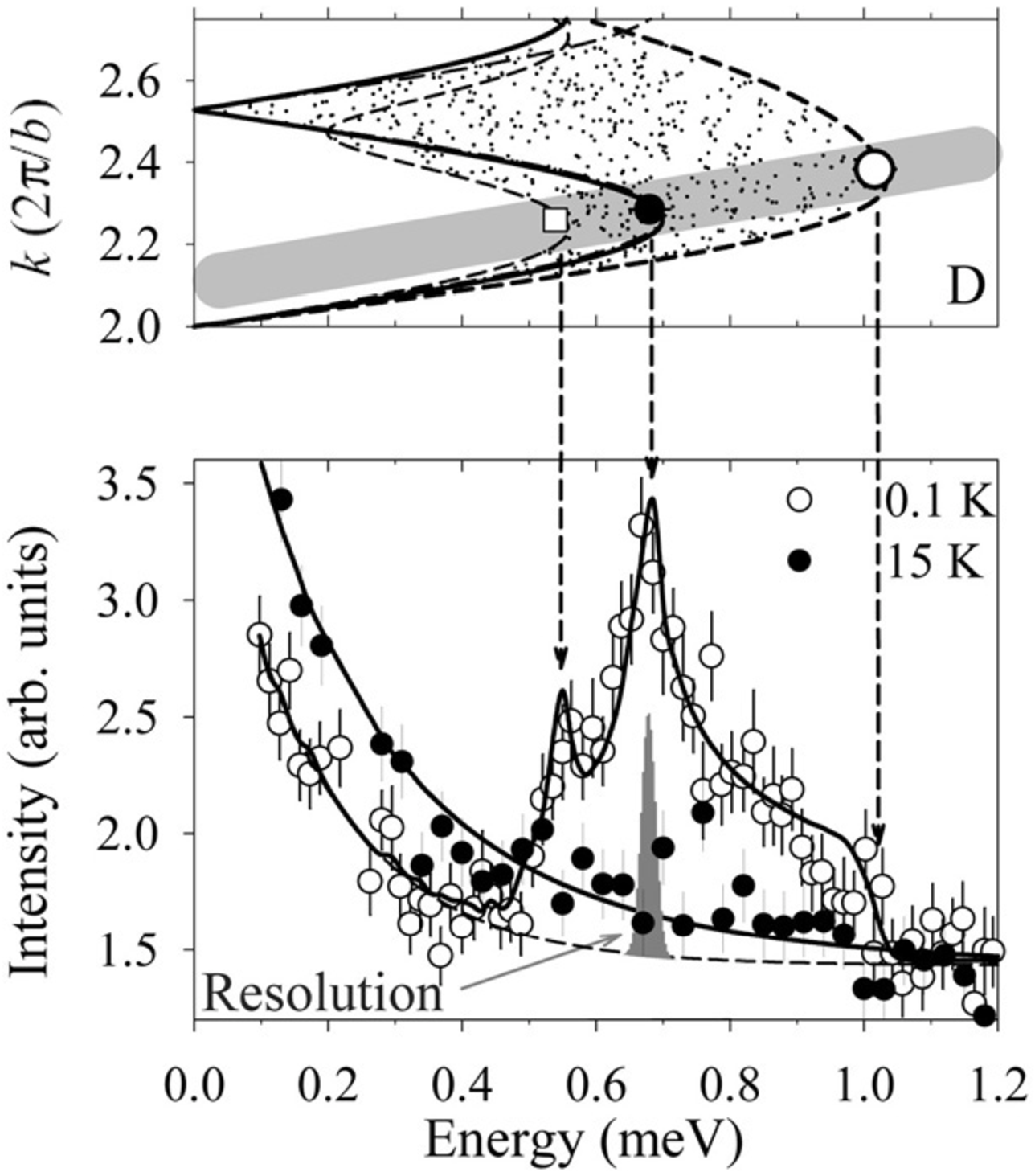}
  {\caption{\label{fig_lineshaped} (Bottom) Intensity
  observed in scan D in the spiral ordered phase at 0.1 K
  (open symbols) and in the paramagnetic phase at 15 K
  (filled circles). Data points are raw neutron counts and
  solid lines are a guide to the eye. The dashed line is
  the estimated non-magnetic background and the grey shaded
  peak indicates the instrumental energy resolution. (Top)
  Intersection of scan D with the dispersion relations.
  Vertical arrows between the top and bottom panels show
  the onset of the magnetic scattering (open square),
  the main peak in the lineshape (solid circle), and the upper
  continuum boundary (open circle), respectively.}}
\end{center}
\end{figure}
Lineshapes of scattering continua are important as they give
information about the underlying quasi-particles and their
interactions. In Cs$_2$CuCl$_4$ the dynamical correlation are
dominated by extended continua as indicated by the dotted areas in
Fig.\ \ref{fig_dispersion_alongkl}.
A representative scan near the antiferromagnetic zone boundary
along $k$
is shown in Fig.\ \ref{fig_lineshaped}(bottom). The scattering is
highly asymmetric with a significant high-energy tail.
Instrumental resolution effects are minimal as the overall extent
of the inelastic scattering is an order of magnitude larger than
the instrumental resolution (grey shaded peak). The non-magnetic
background (dashed line) is modelled by a
constant-plus-exponential function. The magnetic peak disappears
at 15 K (solid circles) and is replaced by a broad, overdamped,
paramagnetic signal. Instead of containing sharp peaks,
characteristic of conventional, $S$=1 magnon excitations, the
observed low-temperature lineshape is dominated by a broad
continuum scattering. To quantify the extent of the magnetic
scattering we define the ``upper boundary" $\omega_{\bm k}^U$ as
the energy above which the observed intensities could not be
distinguished from the non-magnetic background. The upper boundary
dispersion is shown in Fig.\ \ref{fig_dispersion_alongkl} (large
open circles and upper dashed line) and is included to indicate
the region below which most of the magnetic scattering is located.

Representative scans throughout the Brillouin zone are shown in
Fig.\ \ref{fig_irs_scans}, where in this case the data is properly
normalized and corrected for absorption effects, and with the
non-magnetic background subtracted.
Sharp thresholds observed at the lower boundaries of scattering
continua, most clearly seen in scans A and F, are indicative of
highly-coherent excited states, as opposed to the broad,
featureless scattering in systems with spin freezing or random
disorder.

We first analyze the scattering lineshapes in terms of linear
spin-wave theory (LSWT), which provides a good description of the
excitations in unfrustrated 2D square-lattice Heisenberg
antiferromagnets\cite{Hayden91}.

\subsubsection{Lineshape analysis using linear spin-wave theory}
\label{subsubsec_lineshape_lswt}
The spin-wave theory for a frustrated Hamiltonian with a
spiral-ordered ground state (reviewed in Appendix A) predicts a
principal spin-wave mode $\omega_{\bm{k}}$ polarized perpendicular
to the plane of spin rotation (out-of-plane) and two secondary
modes $\omega^+_{\bm{k}}=\omega_{\bm{k}+\bm{Q}}$ and
$\omega^-_{\bm{k}}=\omega_{\bm{k}-\bm{Q}}$, polarized in-plane.
The three dispersions are graphed in Fig.\
\ref{fig_dispersion_alongkl} and the corresponding excitations can
be identified with quantized modes where the total spin component
along the $z$-axis normal to the spiral plane changes by $\Delta
S^z$=0 ($\omega_{\bm k}$), $+1$ ($\omega^{+}_{\bm{k}}$) and $-1$
($\omega^{-}_{\bm{k}}$) ($+z$ defines the sense of rotation in the
spiral).

\vspace{12pt} {\it (a) Sharp magnon excitation at low energies}
\vspace{12pt}

\begin{figure*}[t]
\begin{center}
 \includegraphics[width=17cm,bbllx=1,bblly=224,bburx=555,
  bbury=596,angle=0,clip=]{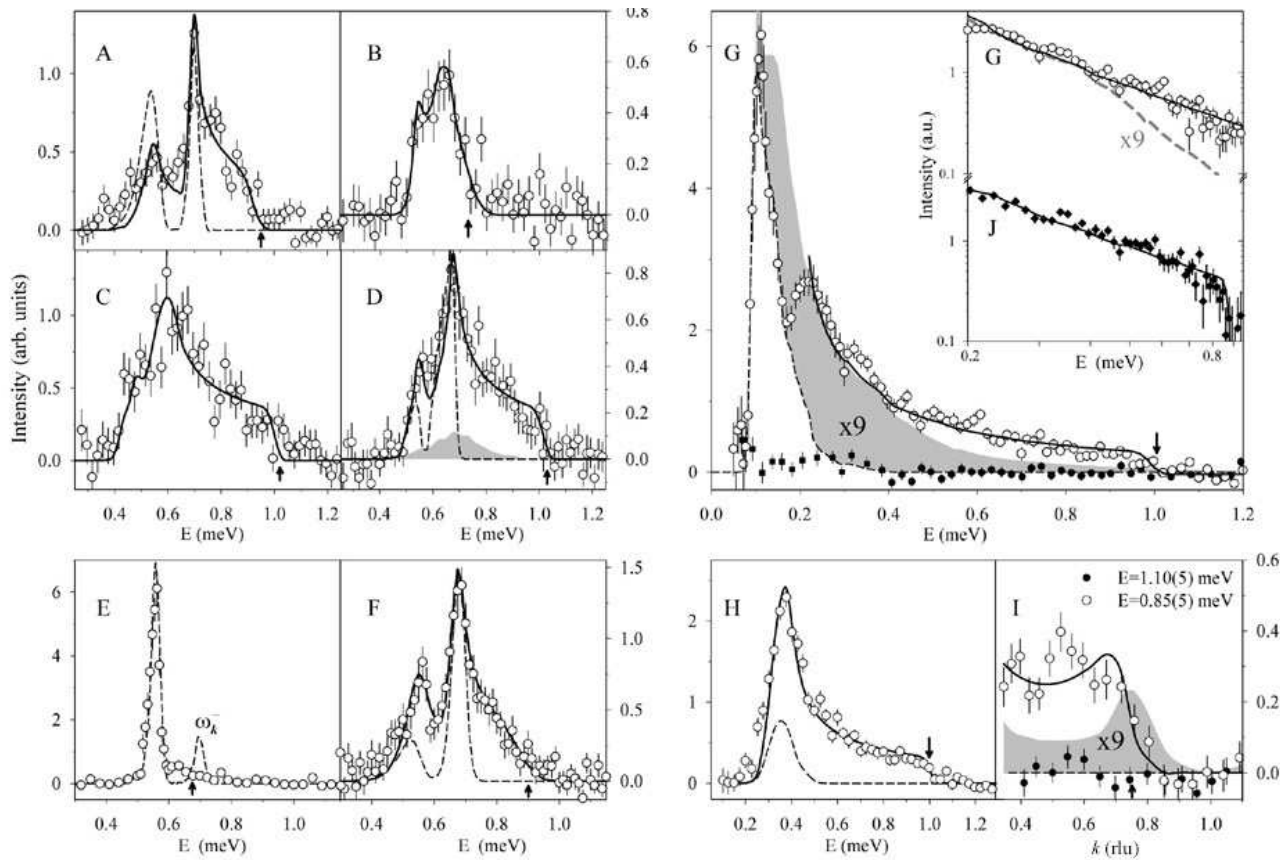}
  {\caption{\label{fig_irs_scans} Magnetic inelastic
  scattering measured in the ordered phase below 0.1 K
  in scans A-J.
  Solid lines are fits to a modified two-spinon cross-section,
  Eq.\ (\ref{eq_lineshape_2spinon_3continua}) for A-D and
  Eq.\ (\ref{eq_lineshape_2spinon}) for G-J. Vertical arrows
  indicate the estimated continuum upper boundary $\omega^U_{\bm k}$.
  Dashed lines show predicted lineshape for polarized
  cycloidal spin waves Eq.\ (\ref{eq_app_1m}) and the dark
  shaded region (D,G,I) indicates the predicted two-magnon
  scattering continuum calculated using Eq.\ (\ref{eq_app_2m})
  and sum rules (scaled up by a factor of 9 in G and I).
  All calculations include the convolution with the spectrometer
  resolution function and the isotropic magnetic form
  factor for Cu$^{2+}$ ions\cite{Wilson95}. Solid points
  in G show the accuracy of the background subtraction for
  wavevectors where no magnetic scattering is expected,
  squares for $k$=0.70(5), $E<$ 0.4 meV, and
  filled circles for $k$=1.00(5), $E>$0.4 meV in Fig.\
  \ref{fig_dispersion_alongkl}(4).}}
\end{center}
\end{figure*}

Scans that probe the low-energy excitations observe a sharp peak
below the onset of the higher-energy continuum scattering. Fig.\
\ref{fig_irs_scans}G shows measurements near the zone-boundary
point along $l$, $\bm{k}$=(0,0.5,1.5) [scan G in Fig.\
\ref{fig_dispersion_alongkl}(3)]. A sharp peak occurs at
$\omega_{\bm k}$=0.107(10) meV followed by strong continuum
scattering with an onset at a slightly higher energy around
$\omega^*\simeq$0.2 meV and extending up to 1 meV. The sharp
low-energy peak can be well described by the spin-wave
cross-section in Eq.\ (\ref{eq_app_1m}) (dashed line), which
predicts a dominant (out-of-plane polarized) magnon at the lowest
energy followed by very much weaker ($\sim 1/6$ intensity)
secondary modes.

\vspace{12pt} {\it (b) Continuum scattering vs. two-magnon
processes} \vspace{12pt}

The largest intensity in scan G is however contained in the
continuum scattering (about 2/3 of the total energy-integrated
intensity).
In contrast, LSWT predicts a much smaller continuum intensity
arising from two-magnon scattering (about 15\%).
In fact the shaded area in Fig.\ \ref{fig_irs_scans}G shows the
two-magnon intensity scaled upwards by a factor of 9 to illustrate
it (the calculation is described in Appendix A and is based on
Eq.\ (\ref{eq_app_2m}) and sum rules for the total scattering).
The two-magnon lineshape would predict the onset of the continuum
scattering immediately above the one-magnon energy $\omega_{\bm
k}$=0.107(10) meV, whereas a distinct separation is observed in
the data with the continuum scattering starting around
$\omega^*\simeq$0.2 meV. The predicted two-magnon intensity drops
off very fast with increasing energy (especially at high energies
$E>$0.6 meV$\simeq \tilde{J}$), whereas the observed scattering
falls off significantly slower [see inset in
\ref{fig_irs_scans}G].

The measured continuum scattering as a function of wavevector at
constant energy (scan I in Fig.\ \ref{fig_dispersion_alongkl}(3))
is compared against the two-magnon cross-section in Fig.\
\ref{fig_irs_scans}I. This energy is beyond the one-magnon cutoff
and only two-magnon processes contribute, and their intensity is
shown by the shaded area(the same $\times$9 scaling factor as for
scan G was used). Apart from largely underestimating the overall
strength of the observed scattering intensity at this energy, the
two-magnon functional form is also very different from the data,
predicting a slightly wider extent in wavevector and intensity
modulations (peak at $k\sim$0.75 originating from an increased
density of states for two-magnon scattering) that are not observed
in the data. Similar disagreement occurs between the observed
continuum scattering and a two-magnon lineshape at other
wavevectors throughout the Brillouin zone. From this analysis we
conclude that the observed functional form (energy- and
wavevector-dependence) of the observed continuum scattering is not
reproduced by a two-magnon cross-section.

The sharp peak observed below the scattering continuum in scans
such as in Fig.\ \ref{fig_irs_scans}G could only be clearly
resolved at low energies below $\sim$ 0.2 meV. With increasing
$\omega_{\bm k}$ this sharp peak appears to gradually merge into
the higher-energy continuum scattering and a typical lineshape is
shown in Fig.\ \ref{fig_irs_scans}H [scan H in Fig.\
\ref{fig_dispersion_alongkl}(3)]. The intensity observed here is
dominated by continuum scattering with a power-law lineshape above
a lower boundary $\omega_{\bm k}$=0.34(2) meV (solid line, Eq.\
(\ref{eq_lineshape_2spinon})) and no separate sharp mode could be
resolved below the continuum lower boundary. A spin-wave
cross-section Eq.\ (\ref{eq_app_1m}) (dashed line) scaled to the
intensity of the one-magnon peak in scan G, significantly
underestimates the intensity observed near the lower boundary of
the continuum scattering. At higher energies the spin-wave
cross-section is also inadequate to describe the low-energy part
of the scattering lineshapes as shown by dashed lines in Fig.\
\ref{fig_irs_scans}A,F.

\vspace{12pt} {\it (c) Polarization of the continuum scattering}
\vspace{12pt}

In principle the polarization of the modes can be extracted from
neutron scattering data because of the directional dependence of
the intensities. We now consider this property to find the
polarization of the scattering with respect to the plane of the
spiral order. We compare the intensities measured at two
equivalent places in the 2D Brillouin zone [A and F in Fig.
\ref{fig_dispersion_surface}(b)], but where the relative weights
of the in-plane and out-of-plane polarizations observed by neutron
scattering are very different. In Fig.\ \ref{fig_irs_scans} scan A
observes mainly in-plane polarized scattering,
whereas scan F is equally sensitive to in-plane- and out-of-plane
polarizations, see Table\ \ref{table_scans}.
Contrary to the polarized spin-wave cross-section (dashed lines,
Eq.\ (\ref{eq_app_1m})), which predicts significant changes in the
relative intensity of the different components of the lineshape
due to the changes in the polarization factors, the two scans
observe very similar lineshapes and the scattering is instead
consistent with an isotropic polarization. The high-energy
continuum scattering is essentially unchanged upon heating above
$T_N$ (see Fig.\ \ref{fig_irs_scans_bc_abovetn}G-I), also
consistent with a mainly {\em isotropic} polarization of the
continuum scattering, independent of the orientation of the spiral
ordering plane.

\vspace{12pt}
{\it (d) Sharp mode near saddle-points}
\vspace{12pt}

Sharp peaks were observed at high energies near special
wavevectors where the 2D dispersion $\omega_{\bm k}$ is at a
``saddle" point as indicated in Fig.\
\ref{fig_dispersion_surface}(a) (such points are located near
$\bm{k}$=(0,$\pm 0.25$,$l$) with $l$ an odd integer and equivalent
positions\cite{sl}). A typical scan is shown in Fig.\
\ref{fig_irs_scans}E. The lineshape here is dominated by a sharp,
resolution-limited peak at $\omega_{(0,-0.23,0.98)}$=0.56(1) meV,
followed by only very weak continuum scattering at higher
energies. This sharp mode is much more intense (by a factor of
4.4(8)) than predicted by extrapolating to those energies the
intensity of the low-energy magnon observed at 0.107(10) meV in
scan G assuming a spin-wave dependence, Eq.\
(\ref{eq_app_1m_outofplane}), suggesting an anomalous intensity
vs. wavevector dependence or different origin (perhaps a
multi-particle bound state) for the saddle-point mode. Also,
spin-wave theory predicts an additional mode near $\omega_{\bm
k}^{-}$=0.7 meV (dashed line), which is however not observed in
the data. The saddle-point mode is consistent with an out-of-plane
polarization by comparing its intensity to that of the isotropic
continuum scattering observed near $k$=0.5.

From this analysis we conclude that LSWT is inadequate to describe
the results, in particular the dominant continuum scattering, and
next we consider another approach based on experience in 1D
systems where LSWT breaks down and highly-dispersive continua are
observed.

\subsubsection{Lineshape analysis using a two-spinon cross-section}
\label{subsubsec_lineshape_2spinon}
Neutrons scatter by changing the total spin by $\Delta S_T =
0,\pm1$ and the observation of dominant scattering continua
indicates that the elementary quasi-particles created in such
processes are not conventional $S$=1 spin waves. A natural
explanation is that quasi-particles have fractional spin quantum
number (spin is a good quantum number since the main
($J,J^{\prime}$) Hamiltonian has isotropic exchanges). By analogy
with the 1D $S$=1/2 HAFC chain [$J^{\prime}=0$] we identify those
quasiparticles with spin-1/2 spinons, created in pairs in a
neutron scattering process. The dominant neutron scattering
cross-section is then a two-spinon continuum, {\it isotropic} in
spin space and extending between dispersive lower and upper
boundaries obtained by convolving two spinon dispersions.

We compare the observed continuum scattering with a generic power-law
lineshape
\begin{equation}
S(\bm{k},\omega)=I_{\bm{k}} \frac{\theta(\omega-\omega_{\bm
k})\theta(\omega^U_{\bm{k}}-\omega)} {[\omega^2-\omega^2_{\bm
k}]^{1-\eta/2}} \label{eq_lineshape_2spinon},
\end{equation}
where $\omega_{\bm{k}}$ and $\omega^U_{\bm k}$ are the
experimentally- determined 2D-dispersive lower and upper
boundaries of the continuum scattering, $I_{\bm k}$ is a
wavevector-dependent intensity factor. This form is
generalized form the two-spinon cross-section in the
1D HAFC chain ($\eta$=1) and has also been proposed
in a theoretical description of 2D frustrated
quantum magnets with deconfined spinons \cite{Chubukov94}.

Scans that probe the continuum scattering over a wide energy range
are shown in Fig.\ \ref{fig_irs_scans}G,J. The intensities
observed above a cross-over energy scale $\omega^*\simeq$0.2 meV
can be well described by a power-law lineshape and fits (solid
lines) to Eq.\ (\ref{eq_lineshape_2spinon}) (assuming lower
boundary fixed by the dispersion $\omega_{\bm k}$) give an
exponent $\eta$=0.74 $\pm$ 0.14 (the inset in Fig.\
\ref{fig_irs_scans}G shows a log plot of the continuum scattering
in scans G and J and the solid lines are power-law fits). The
observed wavevector dependence of the continuum scattering at
constant energy is also consistent with the functional form in
Eq.\ (\ref{eq_lineshape_2spinon}) as shown by the solid line in
Fig.\ \ref{fig_irs_scans}I.

At places close to the zone boundary such as Fig.\
\ref{fig_irs_scans}F the observed lineshape can be described by a
superposition of two continua, each with a power-law lineshape as
in Eq.\ (\ref{eq_lineshape_2spinon}) and onsets at separate lower
boundaries. In the figure the solid line shows a superposition of
two continua with onsets at $\omega^-_{\bm k}$=0.53(2) meV and
$\omega_{\bm k}$=0.67(2) meV. The scattering at $l$=0 shown in
Fig.\ \ref{fig_irs_scans}A-D can be described by a superposition
of three continua with lower boundaries at the dispersion
relations $\omega_{\bm{k}}$, $\omega^-_{\bm{k}}$ and
$\omega^+_{\bm{k}}$, isotropic in spin space and with equal
relative weights (solid lines):
\begin{eqnarray}
S(\bm{k},\omega) &= & I_{\bm{k}} \left[
\frac{\theta(\omega-\omega_{\bm k})}{(\omega^2-\omega^2_{\bm
k})^{1-\eta/2}}  + \frac{\theta(\omega-\omega^-_{\bm
k})}{(\omega^2-(\omega^-_{\bm k})^2)^{1-\eta/2}} \right. \nonumber \\
& + & \left. \frac{\theta(\omega-\omega^+_{\bm
k})}{(\omega^2-(\omega^+_{\bm k})^2)^{1-\eta/2}} \right]
{\theta(\omega^U_{\bm{k}} -\omega)},
\label{eq_lineshape_2spinon_3continua}
\end{eqnarray}
where the exponent $\eta$ has values in the range 0.7 to 1.

To summarize, throughout most of the Brillouin zone and over the
largest energy scales probed the dynamical correlations can be
well-described by continuum lineshapes, characteristic of a
two-spinon cross-section.

\subsection{Dispersion relation and excitation lineshapes
in the 2D spin-liquid phase above T$_N$}
\label{subsec_spinliquid}

\begin{figure}[t]
\begin{center}
  \includegraphics[width=8cm,bbllx=54,bblly=71,bburx=535,
  bbury=715,angle=0,clip=]{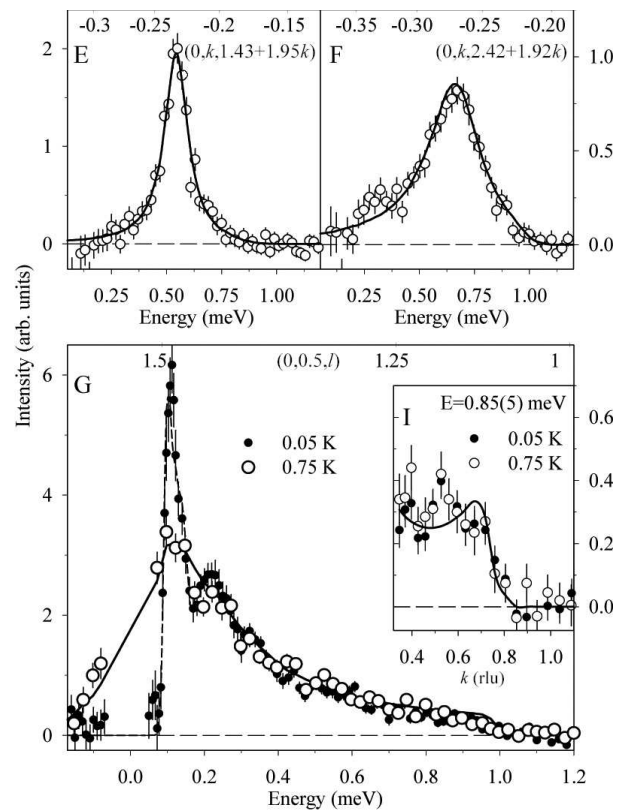}
  {\caption{\label{fig_irs_scans_bc_abovetn} Magnetic inelastic
  scattering measured at 0.75 K in the spin liquid phase above
  $T_N$ (open symbols) and at 0.05 K below $T_N$ (solid symbols).
  Top axis shows wavevector change along scan direction.
  Solid lines in E-F are fits to a damped-harmonic oscillator
  lineshape Eq.\ (\ref{eq_lineshape_dho}) and in G-I are
  guides to the eye (dashed line in G shows a fit to the one-magnon
  cross-section Eq.\ \ref{eq_app_1m}). Data very close to the
  elastic line $E$=0 was omitted in G.}}
\end{center}
\end{figure}

The excitations are thermally broadened which results in a smooth
low-energy tail for the scattering lineshapes as shown in Fig.\
\ref{fig_irs_scans_bc_abovetn}F by data at 0.75 K in the spin
liquid phase above $T_N$. The sharp magnon mode observed at base
temperature below the continuum lower boundary in Fig.\
\ref{fig_irs_scans_bc_abovetn}G(solid points) at $\omega_{\bm
k}$=0.107(10) meV damps out upon heating (open symbols) and merges
with the low-energy tail of the continuum scattering. At high
energies, however, the continuum scattering is practically
unchanged by small temperatures, see Fig.\
\ref{fig_irs_scans_bc_abovetn}I.

\begin{figure}[t]
\begin{center}
 \includegraphics[width=4.5cm,bbllx=215,bblly=227,bburx=420,
  bbury=548,angle=0,clip=]{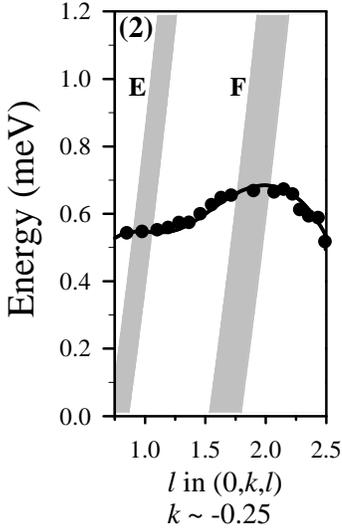}
  {\caption{\label{fig_dispersion_alongl_abovetn} Dispersion
  along $l$ for near-constant $k\sim$-0.25 measured at
  0.75 K in the spin liquid phase above $T_N$. The solid line is
  Eq.\ (\ref{eq_dispersion}) with the same parameters as below
  $T_N$. Intensities observed along scans E-F are shown in
  Fig.\ \ref{fig_irs_scans_bc_abovetn}.}}
\end{center}
\end{figure}

The dispersion relation above $T_N$ maintains a strong 2D
character. The observed incommensuration in the low energy part of
the dispersion relation is plotted as a function of $k$ and $l$ in
Fig.\ \ref{fig_dispersion_alongkl}(5)(open symbols) and is
essentially unchanged compared to base temperatures (solid
symbols). The excitations disperse strongly along both directions
in the 2D plane and the observed dispersion along $l$ for
near-constant $k\sim-0.25$ is plotted in Fig.\
\ref{fig_dispersion_alongl_abovetn}(solid symbols) and is
practically the same as below $T_N$ (solid line). The dispersion
points were extracted by fitting the observed lineshapes to a
damped-harmonic oscillator form (difference of two Lorentzians
centered at $\omega=\pm\omega_{\bm k}$)
\begin{equation}
S(\bm{k},\omega) = I_{\bm{k}}
\frac{\omega\Gamma_{\bm{k}}\omega_{\bm{k}}}
{(\omega^2-\omega^2_{\bm k}-\Gamma_{\bm k}^2)^2 +4\omega^2
\Gamma^2_{\bm k}}
 [n(\omega)+1],
\label{eq_lineshape_dho}
\end{equation}
where $\omega_{\bm k}$ is the dispersion relation, $\Gamma_{\bm
k}$ is the damping rate and $n(\omega)=1/[\exp(\omega/k_BT)-1]$ is
the Bose population factor. This form gave a good description of
the data as shown by typical fits (solid lines) in Fig.\
\ref{fig_irs_scans_bc_abovetn}E-F.

\subsection{Paramagnetic scattering}
\label{subsec_paramagnetic}

\begin{figure}[t]
\begin{center}
  \includegraphics[width=8cm,bbllx=76,bblly=122,bburx=480,
  bbury=669,angle=0,clip=]{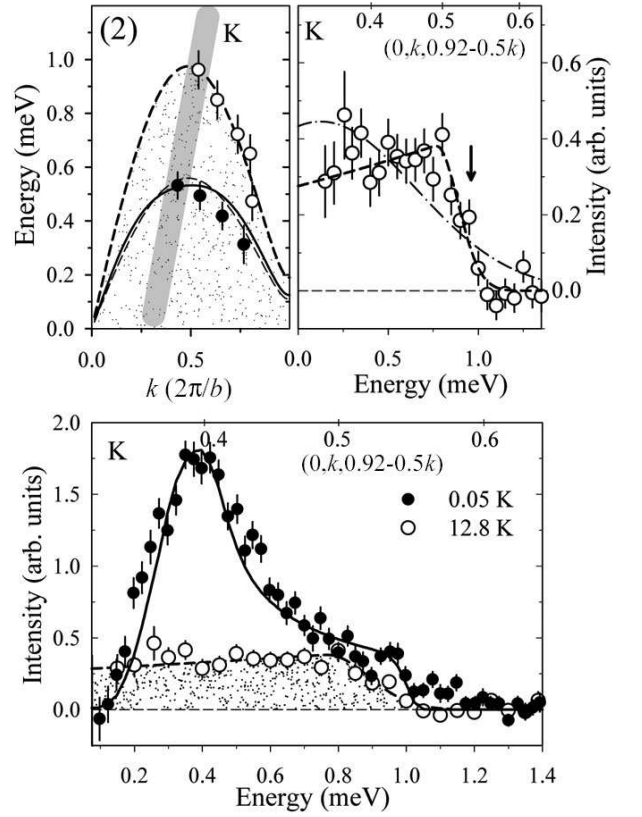}
  {\caption{\label{fig_paramagnetic}(Top left) Extent
  of paramagnetic scattering observed at 12.8 K (dotted
  area). The same wavevector-energy space was probed as for
  the 0.05 K data in Fig.\ \ref{fig_dispersion_alongkl}(2).
  Large open circles are the upper boundary of the
  scattering with the thick (upper) dashed line a guide to the
  eye. Solid points are the square root of the second moment
  of the scattering $\sqrt{\langle\omega^2_{\bm{k}}\rangle}$
  compared to predictions based on sum rules (solid line
  Eq.\ (\ref{eq_secondmoment})); thin dashed line shows the
  same calculation for $T=\infty$.(Top right) Intensity observed
  along scan K. Top axis shows wavevector change along scan
  direction. Thick dashed line is a fit to a plateau-like
  function and dash-dotted line shows a fit to a Gaussian
  with the same second moment. Vertical arrow indicates
  the estimated upper boundary of the scattering.(Bottom)
  Intensity along scan K in the paramagnetic phase (open
  circles 12.8 K) compared to low temperatures (solid circles
  0.05 K). Dotted region highlights the area of the paramagnetic
  scattering. Solid line is a fit of the low-temperature data
  to a power-law lineshape Eq.\ (\ref{eq_lineshape_2spinon}).}}
\end{center}
\end{figure}

Upon further increasing the temperature the lower boundary
collapses and in the paramagnetic phase inelastic intensity is
observed in the form of a broad continuum scattering extending up
to a dispersive upper boundary. Fig.\ \ref{fig_paramagnetic}(top
left) indicates the extent of the scattering (dotted area)
observed at 12.8 K(=2.9 $J$) as a function of wavevector and
energy. The intensity observed in scan K near the
antiferromagnetic zone center along [0$k$0]($k\sim$0.5) where the
scattering has the largest extent in energy is shown in Fig.\
\ref{fig_paramagnetic}(bottom). The paramagnetic lineshape (open
circles) is broad with scattering extending up to nearly the same
upper boundary as the continuum scattering observed at low
temperatures (filled circles).

We quantify the energy width of the scattering in terms of the
normalized second moment $\langle\omega^2_{\bm{k}}\rangle = \int d
\omega \: \omega^2 S(\bm{k},\omega) /\int d\omega \:
S(\bm{k},\omega)$,
which can be directly related to the exchange couplings in the
spin Hamiltonian using sum rules. 
Following Refs.\ \onlinecite{Lovesey87} and
\onlinecite{deGennes58} we obtain for a Heisenberg Hamiltonian to
first order in $\beta=1/k_BT$
\begin{eqnarray}
\langle\omega^2_{\bm{k}}\rangle&=&
\frac{2}{3}S(S+1)\sum_{\bm{\delta}}J_{\bm{\delta}}^2
\left[1-\cos(\bm{k}\cdot\bm{\delta})\right] \times
\nonumber \\
& & \left[1+ \beta
\left(\frac{J_{\bm{\delta}}}{4}+\frac{2}{3}S(S+1)J_{\bm{k}}
\right) \right], \label{eq_secondmoment}
\end{eqnarray}
where the sum is over all exchange couplings $J_{\bm{\delta}}$,
$\bm{\delta}$ is a vector connecting sites and
$J_{\bm{k}}=\frac{1}{2}\sum_{\bm{\delta}} J_{\bm{\delta}}\exp({\rm
i} \bm{k}\cdot \bm{\delta})$ is the Fourier transform of the
exchange couplings. For the 1D $S$=1/2 HAFC model ($J^{\prime}$=0
in Eq.\ (\ref{eq_ham})) the above result predicts that the
paramagnetic scattering at $T=\infty$ has a sinusoidal width
$\sqrt{\langle\omega^2_{\bm{k}}\rangle}=\sqrt{2}J|\sin \pi k|$
that is zero at the ferromagnetic zone center and increases up to
a maximum of $\sqrt{2}J$ at the antiferromagnetic zone center.

The second moment was extracted from scans such as in Fig.\
\ref{fig_paramagnetic}(top right) where the data on the negative
energy side was constructed from the data on the positive energy
side according to the principle of detailed balance. This
procedure gave estimates of $\langle \omega^2_{\bm k}\rangle$
averaged over a small range of $\bm{k}$-values and the results are
plotted in Fig.\ \ref{fig_paramagnetic}(top left) (solid symbols).
The observed second moment near the antiferromagnetic zone center
and its reduction upon approaching the ferromagnetic zone center
are well reproduced by Eq.\ (\ref{eq_secondmoment}) (solid line)
for the 2D Hamiltonian in Eq.\ (\ref{eq_ham}) with the bare
exchange couplings $J=$0.374 meV and $J^{\prime}$=0.128 meV as
determined for Cs$_2$CuCl$_4$ using saturation-field
measurements\cite{Coldea02}. Although this high-temperature method
of probing the Hamiltonian does not give the exchange couplings
$J$ and $J^{\prime}$ directly, it does however provide a measure
of the overall energy scale of the couplings without involving
magnetic fields and is thus an independent consistency check of
the bare exchange couplings and quantum renormalization factors.
The second-moment calculation predicts that no significant changes
are expected upon heating to $T=\infty$ (thin dashed line in Fig.\
\ref{fig_paramagnetic}(top left)) showing that the measurement
temperature (2.9$J$) was sufficiently deep into the paramagnetic
phase.


\section{Discussion} \label{sec_discussion}
\subsubsection{Magnetic excitations and cross-over phase diagram}
\label{subsec_disc_crossover}

Our experiments on Cs$_2$CuCl$_4$ described above observe that in
the spin liquid phase above $T_N$ the dynamical correlations are
dominated by highly-dispersive scattering continua, the hallmark
of deconfined $S$=1/2 spinons of a fractionalized phase. Upon
cooling in the spiral ordered phase below $T_N$ an $S$=1 magnon
with a small scattering weight emerges at low energies as expected
for a phase with a continuous broken symmetry, however the
dominant scattering still occurs in the form of a continuum of
excitations at medium to higher energies and is best described in
terms of pairs of deconfined spinons. This could be understood if
Cs$_2$CuCl$_4$ was in the close proximity of a fractionalized
phase with deconfined spinons, but only weakly perturbed by small
terms in the Hamiltonian. Those weak terms stabilize long-range
magnetic order and create an effective short-range attractive
potential between spinons that is sufficient to create a
two-spinon bound state ($S$=1 magnon) at low energies, but
deconfined spinons still occur at energies higher than the small
scale of the attractive potential. In the spin liquid phase above
$T_N$ small thermal fluctuations overcome this small perturbation
in the Hamiltonian and only deconfined spinons would be expected,
in agreement with experiments which observed two-spinon continua
at all energy scales probed. Proximity of Cs$_2$CuCl$_4$ to a
spinon-binding transition is supported by the observation of sharp
divergencies at the lower boundary of two-spinon continua (most
clearly seen in scans A and F), which indicate a resonant
enhancement of the scattering as precursor to the formation of a
sharp bound state, as noted in Ref.\ \onlinecite{Bernevig01}.
Moreover, evidence that the ordered state is stabilized by only
small terms in the Hamiltonian comes form measurements in in-plane
fields which observe suppression of the spiral order and
transition to a disordered phase at relatively small applied
fields\cite{Coldea01}.

An important question is the nature of the fractionalized phase
that dominates the physics in Cs$_2$CuCl$_4$ and leads to strong
two-spinon continua in the dynamical correlations. Various
approaches have been actively pursued theoretically. A quasi-1D
approach was proposed\cite{Bocquet01}, starting from 1D HAFC
chains along the strongest-exchange direction in the 2D triangular
planes and treating perturbatively all other terms in the
Hamiltonian, including the frustrated coupling $J^{\prime}$; in
this scenario the relevant picture is weakly-coupled
chains\cite{Essler97} where spinons are 1D objects identified with
quantum solitons. However, in Cs$_2$CuCl$_4$ the coupling between
``chains" is relatively large $J^{\prime}=0.34(3)J$ and moreover,
the measured dispersion relations show strong (of order
$J^{\prime}$) 2D modulations at all energy scales, indicating that
the excitations are strongly affected by the two-dimensionality of
the couplings. Explicit 2D approaches to the physics, such as the
resonating-valence-bond (RVB) picture\cite{Anderson73}, have been
considered by other authors\cite{Chung01,Zhou02}. The RVB state
can be phenomenologically described as a spin liquid where spins
are spontaneously paired into singlet bonds that fluctuate
(resonate) between many different configurations to gain quantum
kinetic energy. In this phase spinons are 2D objects with
topological character\cite{Senthil00} and can be physically
described as the two spin-1/2 ends of a broken bond that separate
away through bond rearrangement. Explicit calculations within the
two approaches (quasi-1D or explicit 2D) are required to compare
with the measured dispersion relations and scattering lineshapes
to determine which one best captures the physics in
Cs$_2$CuCl$_4$.

\begin{figure}[t]
\begin{center}
  \includegraphics[width=8cm,bbllx=83,bblly=362,bburx=479,
  bbury=714,angle=0,clip=]{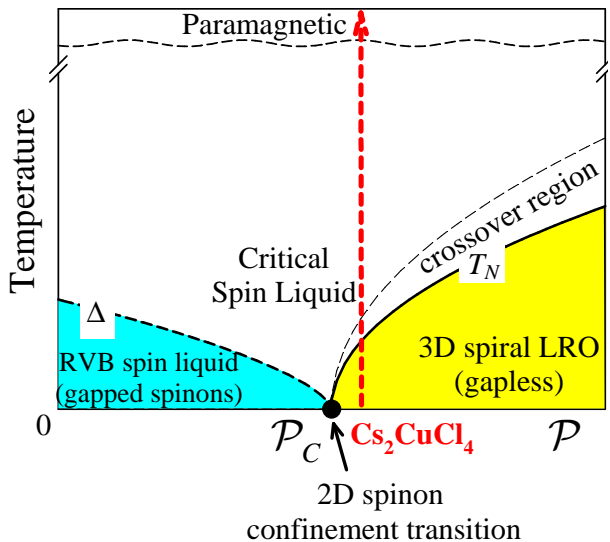}
  {\caption{\label{fig_phase_diagram_schematic}
  (color online) Phase diagram
  of a quasi-2D frustrated quantum magnet with deconfined
  spinons near an instability to spiral long-range order
  driven by a small parameter $\cal{P}$ in the Hamiltonian (such
  as the inter-layer coupling $J^{\prime\prime}$). The vertical
  dashed arrow shows the region that would be probed by
  experiments on Cs$_2$CuCl$_4$.
  Dashed lines mark cross-over to critical behavior from
  gapped fractional spin liquid for $\cal{P}<\cal{P}_C$
  and from renormalized classical on the ordered side
  ($\cal{P}>\cal{P}_C$). At temperatures larger than
  the exchange energies, there is a cross-over
  to paramagnetic behavior (wavy dashed line).}}
\end{center}
\end{figure}

If the fractionalized phase relevant to Cs$_2$CuCl$_4$ is a 2D RVB
phase then a possible phase diagram scenario showing the
transition to incommensurate 3D spiral order driven by some small
parameter in the Hamiltonian (such as the interlayer couplings
$J^{\prime\prime}$) is illustrated in Fig.\
\ref{fig_phase_diagram_schematic}. This phase diagram is inspired
from the proposal\cite{Read91,Chubukov94} that if a 2D frustrated
quantum magnet is driven from a spiral ordered phase into a spin
liquid phase by varying some parameter in the Hamiltonian, then
the spin liquid phase has fluctuating spiral order and deconfined
spin-1/2 spinons. In this scenario the RVB phase is protected by
the gap to break a singlet bond and the transition to order occurs
when the gap closes and an ordered moment can form from condensed
spinons (or pairs of spinons). The magnetic order is an
incommensurate spiral to minimize the frustrated couplings and
symmetry breaking in the spiral plane predicts a gapless Goldstone
spin-wave mode at low energies, which can also be regarded as a
two-spinon bound state stabilized by the weak mean-field effects
of the magnetic order. In this phase diagram scenario deconfined
spinons are expected to occur at higher energies above the scale
of the binding potential, and at all energy scales in the spin
liquid phase above $T_N$ when the mean-field binding effects are
cancelled by small thermal fluctuations, in agreement with
experiments.

Two theories of 2D spin liquid phases with deconfined spinons have
so far been proposed for the main ($J,J^\prime$) Hamiltonian of
Cs$_2$CuCl$_4$, a model with gapped, bosonic spinons\cite{Chung01}
and a model with gapless, fermionic spinons\cite{Zhou02}. In the
latter case the phase diagram in Fig.\
\ref{fig_phase_diagram_schematic} would be modified such that the
fractional spin-liquid phase exists only at the critical point,
i.e. ${\cal P}_C=0$. Another interesting theoretical possibility
is that of a quantum phase with magnetic order and gapless magnons
{\it co-existing} with topological order and gapped spinons
recently discussed in the context of an unfrustrated square
lattice\cite{Lannert02}.

\subsubsection{Paramagnetic scattering}
\label{subsec_disc_paramag}

Neutron scattering studies of the high-temperature paramagnetic
scattering are highly uncommon, whereas in fact they can reveal
important information about the system. Here we consider a
physical picture for the observed scattering in the paramagnetic
phase by making an analogy with two model systems where a number
of exact theoretical results are known. We refer in some detail to
the 1D Haldane-Shastry model (equivalent to an ideal 1D spinon
gas) and the XY chain (equivalent to a non-interacting lattice
fermion gas) where the physics at $T=\infty$ can be understood in
terms of a disordered gas of quasiparticles where as a consequence
of spin conservation rules a neutron scattering process involves
only two-particle states.

The 1D Heisenberg model with $1/r^2$ exchange (Haldane-Shastry
model-HSM)\cite{Haldane93} is equivalent to an ideal spinon gas.
At $T$=0 the ground state is the spinon vacuum and the spin-1
excitations (observable by neutron scattering) are two-spinon
creation precesses\cite{Haldane93}. At $T=\infty$ exact
diagonalizations on finite clusters\cite{Fabricius97} find that
the spin-1 excitations form a continuum in $(k,\omega)$ that
extends up to the same upper boundary $\omega^U_k$ as the
two-spinon continuum at $T$=0; the scattering has a sharp cutoff
at this boundary and no other states at higher energies are
accessible. Combining this result with the fact that the Hilbert
space of the HSM Hamiltonian can be entirely described in a spinon
basis (all eigenstates can be written in terms of states with an
even number of spinons\cite{Haldane93}) suggests that the
dynamical correlations at $T=\infty$ may also be physically
interpreted in terms of two-spinon scattering processes. For
example, $S^{-+}(k,\omega)$ contains scattering events where (i)
two up-spinons are created, (ii) two down-spinons are annihilated
or (iii) an up-spinon is created and a down-spinon is annihilated;
the intensity is a sum of the cross-sections for the above
processes, weighted by the statistical probability of all possible
initial states. Since in the HSM model the spinon dispersion is
{\em not} affected by the presence of other spinons in the ground
state\cite{Haldane93}, the maximum two-spinon energy at $T=\infty$
(when the ground state can be regarded as a dense gas of spinons)
is the same as at $T$=0 (when the ground state is a spinon
vacuum), in agreement with the exact diagonalization
results\cite{Fabricius97} showing an upper boundary unaffected by
temperature.

Another system where only two-particle states contribute to
dynamical correlations (of a conserved operator) is the XY chain
${\cal H}=J \sum_i S^x_iS^x_{i+1} + S^y_iS^y_{i+1}$, which can be
recast into a problem of non-interacting lattice fermions using a
Jordan-Wigner spin-particle mapping (up spin = occupied state,
down spin = empty state). The Hilbert space consists of
eigenstates with a fixed number of particles and the ground state
at $T$=0 has all negative energy states occupied and all positive
energy states empty, where the single-particle dispersion is
$\omega_k=J\cos(2 \pi k)$. The excitations contributing to the
dynamical correlations of the {\em conserved} $S^z$ operator
consist of particle-hole pairs\cite{Katsura70,Fabricius97}:
$S^{zz}(k,\omega)=\sum_{k_1,k_2} (1-f_{k_1}) f_{k_2}
\delta(\omega-\omega_{k_1}+\omega_{k_2}) \delta(k+k_1-k_2)/N$,
where $f_{k}=1/[\exp(\omega_k/k_BT)+1]$ is the statistical
probability that the state $k$ is occupied. At $T=\infty$ every
state is equally probable to be occupied or empty ($f_k=1/2$, the
ground state is a disordered gas of fermions) and the $S^{zz}$
dynamical correlations are a particle-hole pair continuum
contained within $-\omega^U_k \leq \omega \leq \omega^U_k$ with
the same upper boundary as at $T$=0, $\omega^U_k=2J|\sin(\pi k)|$.

The above interpretation of the paramagnetic scattering in the
ideal spinon gas (HSM model) in terms of two-spinon states could
also be extended to the 1D nearest-neighbor HAFC chain. In this
case however interactions between spinons renormalize downwards
the dispersion relation \cite{Muller81}. Therefore the upper
boundary of the two-spinon continuum at $T=\infty$ may be slightly
reduced compared to $T$=0 as a consequence of energy
renormalization effects. This conjecture is consistent with exact
diagonalizations on finite clusters\cite{Fabricius97}, which found
that most of the scattering weight (within a few percent) at
$T=\infty$ is located below the $T$=0 two-spinon boundary; the
same study\cite{Fabricius97} also found that at $T=\infty$ the
lineshapes near the antiferromagnetic zone center approached a
distinctly non-Gaussian shape, better approximated by a
plateau-like function centered at zero energy (similar in shape to
the solid curve in Fig.\ \ref{fig_irs_scans_bc_abovetn}I but where
the horizontal axis is energy). The lineshape observed in Fig.\
\ref{fig_paramagnetic}(top right) is better described by a
plateau-like function (thick dashed line) than by a Gaussian of
the same second moment (dash-dotted line). In this comparison both
curves are centered at $\omega$=0 and are multiplied by
$(\omega/k_BT)/[1-\exp(-\omega/k_BT)]$ to account for thermal
population effects to first order in $\omega/k_BT$.

The observation of continuum scattering extending up to an upper
boundary that is only slightly reduced compared to the
low-temperature extent of the continuum scattering [see Fig.\
\ref{fig_paramagnetic}(bottom)] suggests that the paramagnetic
lineshapes in Cs$_2$CuCl$_4$ may be understood in terms of
(mainly) two-spinon scattering processes.

\section{Conclusions}
\label{sec_conclusions}
In conclusion we have measured the magnetic excitations in the
quasi-2D frustrated quantum antiferromagnet Cs$_{2}$CuCl$_{4}$
using high-resolution time-of-flight neutron spectroscopy. In the
2D spin liquid phase above the magnetic ordering transition the
dynamical correlations are dominated by highly-dispersive
scattering continua, characteristic of fractionalization of spin
waves into pairs of deconfined, $S$=1/2 spinons. Boundaries of the
scattering continua are strongly-2D dispersive indicating that 2D
effects are important up to all energy scales of the excitations.
Upon cooling into the 3D spiral ordered phase sharp magnon modes
emerge at the lowest energies, but the dispersion relations and
continuum scattering at medium to high energies were essentially
unchanged compared to above $T_N$. This suggests a dimensional
crossover from 3D $S$=1 spin waves at low energies to 2D $S$=1/2
spinons at medium to high energies. A phase diagram scenario
locating Cs$_2$CuCl$_4$ in the proximity of a spinon
de-confinement transition was proposed.

We hope these results will stimulate further theoretical work in
the field of frustrated quantum magnets and fractionalized phases;
knowing the Hamiltonian rigorously\cite{Coldea02} should provide a
test of the various theoretical approaches. We also note that
crucial in our experimental exploration of the physics and in
building up a phenomenological picture of the excitations was the
use of time-of-flight neutron spectroscopy, which allowed probing
the dynamical correlations over a very large part of phase space.

\section{Acknowledgements}
\label{sec_acknowledgements}
We especially wish to thank A.M. Tsvelik for numerous insightful
conversations. We also acknowledge technical support from M.
Telling and M.A. Adams and wish to thank J.B. Marston, T. Senthil,
S. Sachdev, S.L. Sondhi, R. Moessner, F.H.L. Essler, X.G. Wen,
R.A. Cowley and M. Bocquet for many fruitful discussions and
interest in the work. ORNL is managed for the US DOE by
UT-Battelle, LLC, under contract DE-AC05-00OR22725.

\appendix
\section{Linear spin-wave theory for a spiral-ordered magnet}
\label{app_spinwaves}
In this section we describe the ground state and excitations of
the frustrated Hamiltonian on the anisotropic triangular lattice
in Fig.\ \ref{fig_cs2cucl4_structure}(b) using linear spin-wave
theory. This approach assumes that spins order in a well-defined
structure (found by minimizing the mean-field energy) and treats
excitations as local spin deviations that propagate by inter-site
hopping. Linearizing the equations of motion results in wave-like
propagating modes, the magnon quasi-particles, which have
quantized spin angular momentum $S$=1 and Bose statistics. We
discuss the dynamical correlations parameterized in terms of one-
and two-magnon scattering processes.

\begin{figure}[t]
\begin{center}
   \includegraphics[width=8cm,bbllx=187,bblly=310,bburx=559,
   bbury=720,angle=0,clip=]{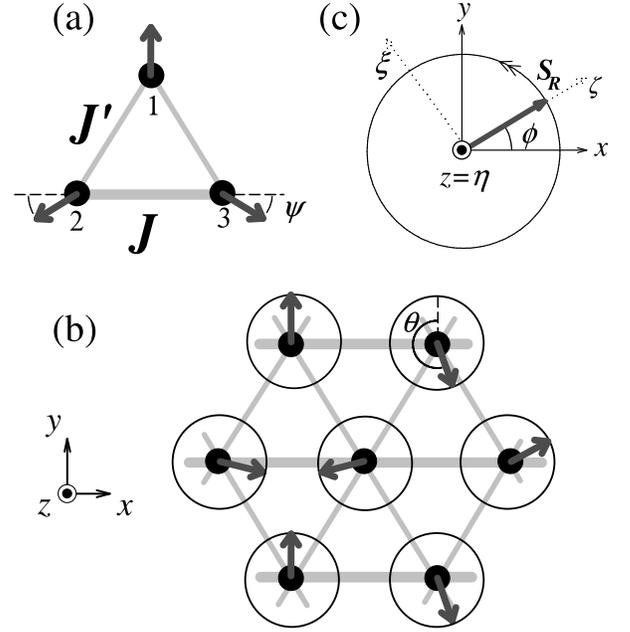}
   {\caption{\label{fig_spin_rotation}
   (a) Schematic diagram of the mean-field ground state for three spins
   located in the corners of an isosceles triangle with antiferromagnetic
   couplings $J$ (horizontal thick bond) and $J^{\prime}$ (thin zig-zag bonds).
   (b) Spiral order on the anisotropic triangular lattice obtained
   by generalizing the picture shown in (a). Thick arrows indicate the
   ordered spin direction, which rotates in the ($x,y$) plane by an
   incommensurate angle $\theta=2\pi Q$ between neighboring sites
   in the direction of the strong exchange $J$ (horizontal thick
   bonds). (c) Definition of the rotating reference frame
   ($\xi$,$\eta$,$\zeta$) where $\zeta$ is along the local mean spin
   direction and $\eta$ is perpendicular to the rotation plane.}}
\end{center}
\end{figure}

To find the mean-field ground state it is illuminating to consider
first an isosceles triangle [see Fig.\
\ref{fig_spin_rotation}(a)], which is the building block of the
full two-dimensional anisotropic triangular lattice in Fig.\
\ref{fig_cs2cucl4_structure}(b). In the absence of the zig-zag
coupling $J^{\prime}$=0 the strong antiferromagnetic exchange $J$
favors antiparallel alignment of the spins 2 and 3. This collinear
order becomes unstable in the presence of a coupling $J^{\prime}$
and the mean field created by spin 1 induces a spin-flop-like tilt
of the other two spins as indicated in the figure. The angular
tilt $\psi$ depends on the relative strength $J^{\prime}/J$ and is
therefore incommensurate with the lattice. Generalization of this
model to the full 2D lattice results in the non-collinear spiral
order indicated in Fig.\ \ref{fig_spin_rotation}(b).
The ordered spin at site $\bm{R}$ is $\langle \bm{S}_{\bm
R}\rangle=\langle S \rangle \cos(\bm{Q}\cdot\bm{R}+\phi_0)
\hat{\bm{x}} + \langle S \rangle \sin(\bm{Q}\cdot\bm{R}+\phi_0)
\hat{\bm{y}}$, where $\langle S \rangle $ is the magnitude of the
ordered spin moment and $\phi_0$ is an arbitrary phase angle.
$\hat{\bm{x}}$ and $\hat{\bm{y}}$ are orthogonal unit vectors
defining the plane of spin rotation. The ordering wavevector
$\bm{Q}$ is found by minimizing the exchange energy (per spin)
$J_{\bm{k}}=J\cos (2\pi k) +2 J^{\prime}\cos (\pi k )\cos (\pi
l)$, $\bm{k}=(h,k,l)$ and is ${\bm Q}=(0,0.5+\epsilon_c,0)$ where
$\epsilon_c=\sin^{-1}[J^{\prime}/(2J)]/\pi$ is the
incommensuration relative to N\'{e}el order. In the mean-field
approximation the incommensurate spiral order occurs immediately
as the chains along $J$ become coupled, i.e. $J^{\prime}\neq$0,
and is stable throughout the range $0<J^{\prime}/J<2$, which
includes the isotropic triangular lattice $J^{\prime}/J$=1 for
which $\epsilon_c=1/6$.

An incommensurate spiral order as depicted in Fig.\
\ref{fig_spin_rotation}(b) is observed\cite{Coldea96} in
Cs$_2$CuCl$_4$ at temperatures below $T_N$=0.62(1) K. The ordered
spins rotate in cycloids, nearly contained in the ($bc$) plane due
to a small anisotropy. The angle between consecutive spins along
$b$ is $\theta=2\pi Q$=180$^{\circ}$+10.8(8)$^{\circ}$ which give
an incommensuration $\epsilon_0$=0.030(2). The measured
incommensuration is significantly smaller than the classical value
$\epsilon_c$=0.054 corresponding to the measured exchange
parameters\cite{Coldea02} $J^{\prime}/J$=0.34(3). This large
renormalization $\epsilon_0/\epsilon_c$=0.56 is due to quantum
fluctuations not captured by the classical mean-field
approximation, but well reproduced by calculations which include
nearest-neighbor spin-singlet correlations in the incommensurate
ground state\cite{Weihong99,Chung01}. Those correlations favor
antiferromagnetic alignment of the paired spins, therefore acting
to reduce the incommensuration as observed.

Since the ordered spin direction rotates as one moves from site to
site it is convenient to define a local reference frame
($\xi,\eta,\zeta$) such that the mean spin direction at each site
appears along the $\zeta$ axis for all sites. This transformation
is illustrated in Fig.\ \ref{fig_spin_rotation}(c) and is defined
by
\begin{eqnarray}
S^x & = & S^{\zeta}\cos \phi - S^{\xi} \sin \phi, \nonumber \\
S^y & = & S^{\zeta}\sin \phi + S^{\xi} \cos \phi, \nonumber \\
S^z & = & S^{\eta}, \label{eq_app_spin_rotation}
\end{eqnarray}
where the rotation angle at site $\bm R$ is
$\phi=\bm{Q}\cdot\bm{R}+\phi_0$. Spin deviation operators are
expressed in terms of magnon Bose operators using the
Holstein-Primakoff \cite{Holstein40} formalism
\begin{eqnarray}
S^{+}_{\bm{R}} & =& S^{\xi}_{\bm{R}} + {\rm i} S^{\eta}_{\bm{R}} = \sqrt{2S} a_{\bm R} + \ldots , \nonumber \\
S^{-}_{\bm{R}} & =& S^{\xi}_{\bm{R}} - {\rm i} S^{\eta}_{\bm{R}} = \sqrt{2S} a^{\dag}_{\bm R} + \ldots, \nonumber \\
S^{\zeta}_{\bm R} & =& S  - a^{\dag}_{\bm{R}}a_{\bm{R}},
\label{eq_app_holstein}
\end{eqnarray}
where $\ldots$ stand for higher-order terms involving three or
more operators. $a^{\dag}_{\bm{R}}$ creates a spin deviation at
site $\bm{R}$ and its Fourier transform $a^{\dag}_{\bm
k}=\sum_{\bm R} a^{\dag}_{\bm R}e^{-{\rm i} \bm{k} \cdot
\bm{R}}/\sqrt{N}$ creates an extended wave ($N$ is the total
number of spins).

In a $1/S$ expansion the leading term in the Hamiltonian becomes a
quadratic form of magnon operators. This can be diagonalized using
standard techniques to obtain the wave-vector-dependent spin-wave
energies\cite{Nagamiya67}
\begin{equation}
\omega_{\bm k}= 2S \sqrt{\left(J_{\bm k}-J_{\bm Q}\right)
               \left[\left(J_{\bm{k}-\bm{Q}} + J_{\bm{k} +\bm{Q}} \right)/2-J_{\bm
               Q}\right]},
               \label{eq_app_dispersion}
\end{equation}
where $J_{\bm k}=J\cos(2\pi k)+2J^{\prime}\cos(\pi k ) \cos(\pi
l)$ is the Fourier transform of the exchange couplings.
Higher-order terms in the Hamiltonian amount to interactions
between magnons, which are expected to produce (relatively) small
renormalizations of the above spin-wave energies.
A typical plot of the dispersion relation in Eq.\
(\ref{eq_app_dispersion}) is shown in Fig.\
\ref{fig_dispersion_surface}(a). The dispersion has a sinusoidal
shape with gapless modes at the ferromagnetic zone center
$\bm{k}=\bm{0}$ and at the incommensurate ordering points
$\bm{k}=\pm\bm{Q}$ in the 2D Brillouin zone. The $\omega_{\bm{0}}$
mode corresponds to a uniform rigid rotation of all spins in the
ordering plane and is called a phason, whereas the
$\omega_{\pm\bm{Q}}$ modes correspond to fluctuations where the
plane in which spins rotate cants away from the ordering plane.
Those three gapless excitations are Goldstone modes associated
with the breaking of symmetry in spin space by the spiral
long-range order [in the presence of a small easy-plane anisotropy
the mode $\omega_{\pm\bm{Q}}$ acquires a gap but $\omega_{\bm{0}}$
remains gapless as it involves spin rotations inside the
easy-plane].

Information about the nature and properties of the spin
excitations is contained in the dynamical correlation function
$S^{\alpha\beta}(\bm{k},\omega)$, directly measured by neutron
scattering \cite{Lovesey87}. This is defined as the Fourier
transform of the space and time correlation function
\begin{equation}
S^{\alpha\beta}(\bm{k},\omega)=\frac{1}{2\pi\hbar}\int_{-\infty}^{+\infty}
dt \sum_{\bm{R}} \langle S^{\alpha}_{\bm 0}(0) S^{\beta}_{\bm
R}(t) \rangle e^{i (\omega t-\bm{k} \cdot \bm{R})},
\end{equation}
where $\alpha$ and $\beta$ label Cartesian axes and $\langle \dots
\rangle$ denotes a ground state average. Because the Hamiltonian
(\ref{eq_ham}) conserves the total spin component $S^z_T$, the
off-diagonal terms cancel out $S^{\alpha\beta}(\bm{k},\omega)=0$
for $\alpha \neq \beta$, see Ref. \onlinecite{Lovesey87}.

One-magnon ($1\cal{M}$) excitations occur in the fluctuations
polarized transverse to the ordered spin direction $\zeta$. The
transverse dynamical correlations at $T$=0 are obtained as
\cite{Nagamiya67}
\begin{eqnarray}
S^{\xi\xi}_{1\cal{M}}(\bm{k},\omega) & \!\!\!= \!\!\!&
\frac{\tilde{S}}{2}
\left|u_{\bm k}+v_{\bm k}\right|^2
\delta(\omega-\omega_{\bm
k}), \nonumber \\
S^{\eta\eta}_{1\cal{M}}(\bm{k},\omega) & \!\!\!=\!\!\!&
\frac{\tilde{S}}{2}
\left| u_{\bm k}-v_{\bm k}\right|^2 \delta(\omega-\omega_{\bm k}).
\label{eq_app_1m_correlations}
\end{eqnarray}
Sharp one-magnon peaks occur at wavevectors and energies given by
the dispersion relation $\omega_{\bm k}$. The scattering
intensities depend on the functions $u_{\bm k}$ and $v_{\bm k }$
defined as $u_{\bm k}=\cosh \theta_{\bm k}$ and $v_{\bm k }=\sinh
\theta_{\bm k}$ where $\tanh{2 \theta_{\bm k}}=B_{\bm k}/A_{\bm
k}$ and
\begin{eqnarray}
A_{\bm k}& =& 2S \left\{ \frac{J_{\bm k}}{2}+ \frac{1}{4}\left[
J_{\bm{k} - \bm{Q}} + J_{\bm{k} + \bm{Q}} \right]  -J_{\bm Q} \right\}, \nonumber \\
B_{\bm k}& =& 2S \left\{ \frac{J_{\bm k}}{2} - \frac{1}{4}\left[
J_{\bm{k} - \bm{Q}} + J_{\bm{k} + \bm{Q}} \right] \right\}
.\nonumber
\end{eqnarray}
The dispersion relation in Eq.\ (\ref{eq_app_dispersion}) can
alternatively be expressed as
$\omega_{\bm{k}}=\sqrt{A^2_{\bm{k}}-B^2_{\bm k}}$.

\begin{figure}[t]
\begin{center}
   \includegraphics[width=7cm,bbllx=139,bblly=293,bburx=445,
   bbury=738,angle=0,clip=]{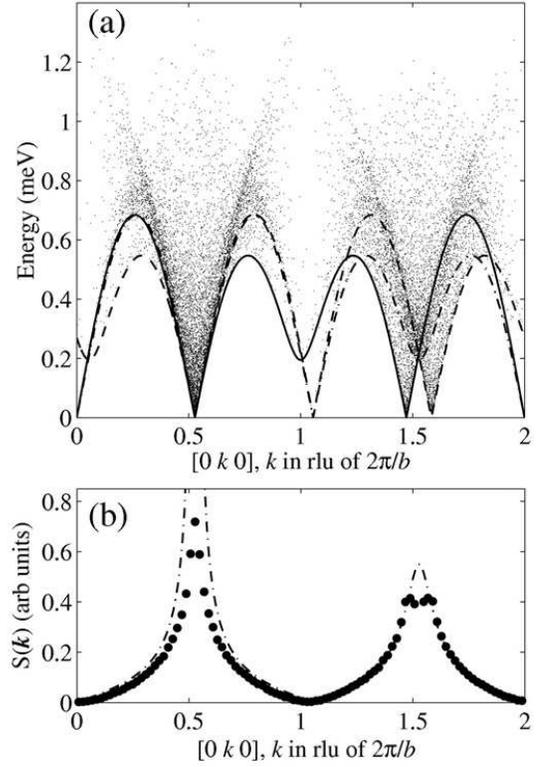}
   {\caption{\label{fig_1and2m_kmq}
   (a) Two-magnon scattering intensity
   $S^{\zeta\zeta}_{2\cal{M}}(\bm{k}-\bm{Q},\omega)$
   in Eq.\ (\ref{eq_app_2m_inplane}) as a function of energy and
   wave vector along $\bm{b}^*$. Density of scattered points
   represents intensity. The plotted points are random
   two-magnon scattering events generated using a
   Monte Carlo algorithm for the sum in Eq.\ (\ref{eq_app_2m_correlations})
   as described in the text. The displayed two-magnon scattering
   continuum is bounded at low energies by the lower of the three
   dispersion relations $\omega_{\bm{k}}$ (solid line),
   $\omega_{\bm{k}-2\bm{Q}}$ (dashed line) and $\omega_{\bm{k}-\bm{Q}}$
   (dash-dotted line).
   (b) Energy-integrated two-magnon scattering intensity as a function
   of wave vector along $\bm{b}^{*}$. Solid points represent
   $\int d\omega S^{\zeta\zeta}_{2\cal{M}}(\bm{k}-\bm{Q},\omega)$
   calculated using a distribution of scattering events such as
   shown in (a). For comparison, the intensity of the in-plane
   magnon with dispersion $\omega_{\bm{k}-\bm{Q}}$ is shown by
   the dash-dotted line
   ($\int d\omega S^{\xi\xi}_{1\cal{M}}(\bm{k}-\bm{Q},\omega)$
   see Eq.\ (\ref{eq_app_1m_inplane})).
   Calculations are plotted for renormalized couplings
   $\tilde{J}$=0.61 meV and $\tilde{J^{\prime}}/\tilde{J}$=0.175
   that fit the observed dispersion relation 
   in Cs$_2$CuCl$_4$ plotted in Fig.\ \ref{fig_dispersion_alongkl}.
   One- and two-magnon scattering intensities were normalized
   against sum rules as described in the text with
   $\Delta S/S\sim$0.25 estimated from experiments
   \cite{Coldea96}; the term $\langle uv \rangle^2$ was calculated
   to be very small and was neglected.}}
\end{center}
\end{figure}

Fluctuations in the longitudinal spin component ($\parallel
\zeta$) are expressed as a product of two local magnon operators
in Eq.\ (\ref{eq_app_holstein}). After transformation to normal
operators that diagonalize the quadratic Hamiltonian, the
longitudinal dynamical correlation function is expressed in terms
of two-magnon ($2\cal{M}$) scattering processes. At $T$=0 only
two-magnon creation processes can occur and their cross-section is
given by \cite{Heilmann81}
\begin{eqnarray}
S^{\zeta\zeta}_{2\cal{M}}(\bm{k},\omega) &\!\!\! =& \!\!\!
\frac{1}{2N}\!\!\sum_{\bm{k}_1,\bm{k}_2} \!\!f(\bm{k}_1,\bm{k}_2)
\times \nonumber \\
& & \!\!\!\!\!\!\!\!\!\!\!\!\!\!\!\!\!\!\!\!\!\!\! \times
\delta(\omega-\omega_{\bm{k}_1}-\omega_{\bm{k}_2})
\delta(\bm{k}+\bm{k}_1-\bm{k}_2 +\bm{\tau}),
\label{eq_app_2m_correlations}
\end{eqnarray}
where
$f(\bm{k}_1,\bm{k}_2)=\left|u_{-\bm{k}_1}v_{\bm{k}_2}+u_{\bm{k}_2}v_{-\bm{k}_1}\right|^2$
is the scattering weight. This formula is valid in the harmonic
approximation where magnon-magnon interactions are neglected, i.e.
only the quadratic part is retained in the Hamiltonian expansion
in terms of magnon operators. In Eq.\
(\ref{eq_app_2m_correlations}) $\bm{k}_1$ and $\bm{k}_2$ are the
wave vectors of the two magnons, which can be created anywhere in
the first Brillouin zone and $\bm{\tau}$ is a vector of the
reciprocal lattice. For a given wavevector transfer $\bm k$
two-magnon processes contribute an extended scattering continuum
at energies above a lower threshold. This threshold corresponds to
creating one magnon in the lowest energy state (in this case
$\omega_{\bm{k}_1}$=0 at a wave vector $\bm{k}_1=\bm{0},-\bm{Q}$
or $+\bm{Q}$) and the other magnon at another place in the zone.
The sum in (\ref{eq_app_2m_correlations}) was calculated
numerically using a weighted Monte Carlo method described in Ref.
\onlinecite{Tennant95b}. From a large number of randomly-generated
two-magnon scattering processes a distribution of $2\times10^6$
events was selected according to a probability proportional to the
scattering weight $f(\bm{k}_1,\bm{k}_2)$; this gave the complete
picture of the two-magnon intensity in the whole Brillouin zone.
Fig.\ \ref{fig_1and2m_kmq}(a) shows a plot of the resulting
intensity shifted in wave vector by $\bm{Q}$ to obtain the
quantity $S^{\zeta\zeta}_{2\cal{M}}(\bm{k}-\bm{Q},\omega)$
relevant for comparison with experiments. The intensities obtained
after numeric summation were then normalized to satisfy the sum
rule for the total two-magnon scattering in the Brillouin zone, as
discussed below.

Longitudinal inelastic processes described in Eq.\
(\ref{eq_app_2m_correlations}) can occur because the spin moment
is not fully ordered in the ground state. This disordering is due
to zero-point quantum fluctuations and the spin reduction
calculated in the linear spin-wave approximation is $\Delta
S=\sum_{\bm{k}}|v_{\bm{k}}|^2/N$ where $\bm k$ spans the Brillouin
zone. Applied to the relevant 2D couplings in Cs$_2$CuCl$_4$
($J^{\prime}/J$=0.34(3)) this gives a very large spin reduction
$\Delta S$=0.43, see Ref.\onlinecite{Merino99}, however mean-field
effects from the inter-layer couplings and anisotropy terms not
included in this calculation can quench some of the low-energy
fluctuations and partially restore order, leading to a smaller
$\Delta S$. The total two-magnon scattering intensity integrated
over energy and wavevector in a Brillouin zone is $\Delta S
(1+\Delta S) + \langle uv \rangle^2$ (per spin), where $\langle uv
\rangle=\sum_{\bm k} u_{\bm k} v_{\bm k}/N$ is in general non zero
if the ordering wavevector $\bm Q$ is incommensurate. Zero-point
fluctuations also affect the one-magnon scattering by reducing the
intensity factor in Eq.\ (\ref{eq_app_1m_correlations}) from $S$
to $\tilde{S}=S-\Delta S-\langle uv\rangle^2(1+2\Delta S)^{-1}$.
With this normalization the sum rule for the total scattering in
the Brillouin zone $S(S+1)$ is exhausted by elastic Bragg
scattering $(S-\Delta S)^2$ shared between the two positions
$\bm{k}=\pm\bm{Q}$, inelastic one-magnon $\tilde{S}(1+2\Delta S)$
and two-magnon processes. Higher-order scattering processes
involving three or more magnons are neglected here since their
scattering weight decreases very rapidly with increasing particle
number. Those processes redistribute some of the scattering weight
in the zone towards higher energies.

To obtain the dynamical correlations measured in an experiment,
Eqs.\ (\ref{eq_app_1m_correlations}) and
(\ref{eq_app_2m_correlations}) are converted to the fixed
laboratory reference frame ($x,y,z$) using the transformation in
Eq.\ (\ref{eq_app_spin_rotation}). Sharp one-magnon peaks occur
both in the out-of-plane ($\parallel z$)
\begin{equation}
S^{zz}_{1\cal{M}}(\bm{k},\omega) =
S^{\eta\eta}_{1\cal{M}}(\bm{k},\omega)=
\frac{\tilde{S}}{2}\frac{A_{\bm k}-B_{\bm k}}{\omega}
\delta(\omega -\omega_{\bm k}), \label{eq_app_1m_outofplane}
\end{equation}
as well as in the in-plane ($\parallel x,y$) dynamical
correlations
\begin{eqnarray}
S^{xx}_{1\cal{M}}(\bm{k},\omega) & = &
S^{yy}_{1\cal{M}}(\bm{k},\omega)\nonumber\\
& = & \frac{1}{4}\left[
S^{\xi\xi}_{1\cal{M}}(\bm{k}-\bm{Q},\omega)
+S^{\xi\xi}_{1\cal{M}}(\bm{k}+\bm{Q},\omega)\right] \nonumber \\
& & \nonumber \\
& = &
\frac{\tilde{S}}{8}\frac{A_{\bm{k}-\bm{Q}}+B_{\bm{k}-\bm{Q}}}
{\omega} \delta(\omega -\omega_{\bm{k}-\bm{Q}})
+ \nonumber \\
& & \nonumber \\
& & \frac{\tilde{S}}{8} \frac{A_{\bm{k}+\bm{Q}}+B_{\bm{k}+\bm{Q}}}
{\omega} \delta(\omega -\omega_{\bm{k}+\bm{Q}})
\label{eq_app_1m_inplane}
\end{eqnarray}
In total three spin-wave modes are observed: a principal mode
$\omega_{\bm k}$ polarized out-of-plane ($\parallel z$) and two
secondary modes $\omega^-_{\bm{k}}=\omega_{\bm{k}-\bm{Q}}$ and
$\omega^+_{\bm{k}}=\omega_{\bm{k}+\bm{Q}}$ polarized in-plane
($\parallel x,y$). Those three modes can also be regarded as
excitations with quantized spin moment along the $z$-axis normal
to the spiral plane, i.e. the principal mode $\omega_{\bm k}$ has
spin $S^z$=0 and the two secondary modes $\omega^{\pm}_{\bm k}$
have spin $S^z=\pm 1$ ($+z$ defines the sense of rotation in the
spiral). The three dispersions are plotted along the $\bm{b}^*$
and $\bm{c}^*$ directions in Fig.\ \ref{fig_dispersion_alongkl}
and for all modes the intensity decreases as $1/\omega$ with
increasing energy. The secondary modes are images of the main mode
displaced in wavevector by $-\bm{Q}$ and $+\bm{Q}$. This apparent
splitting into three modes is a characteristic feature of systems
with helical order, i.e. helix, cycloid, and cone
\cite{Nagamiya67}.


\begin{figure}[t]
\begin{center}
   \includegraphics[width=6cm,bbllx=1,bblly=1,bburx=357,
   bbury=360,angle=0,clip=]{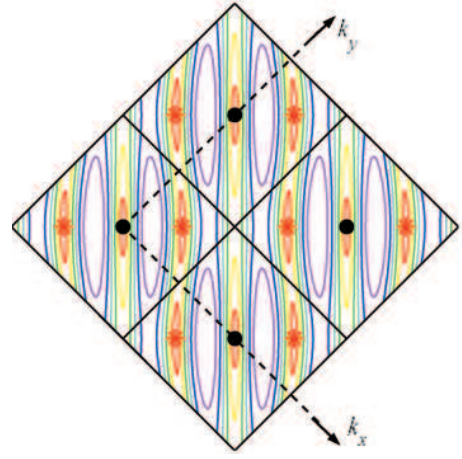}
   {\caption{\label{fig_sl}
   (color online) Contour plot of the 2D
   dispersion relation using a square-lattice
   representation of Fig.\ \ref{fig_cs2cucl4_structure}(b) ($J^{\prime}$
   along the square edges). Filled circles are zone centers and
   stars show positions of incommensurate magnetic Bragg peaks.}}
\end{center}
\end{figure}

The two-magnon scattering is polarized in-plane and is obtained
using transformation (\ref{eq_app_spin_rotation}) as
\begin{eqnarray}
S^{xx}_{2\cal{M}}(\bm{k},\omega) & = &
S^{yy}_{2\cal{M}}(\bm{k},\omega) \nonumber \\
& = & \frac{1}{4}\left[
S^{\zeta\zeta}_{2\cal{M}}(\bm{k}-\bm{Q},\omega)+S^{\zeta\zeta}_{2\cal{M}}(\bm{k}+\bm{Q},\omega)\right].
\nonumber \\ & &  \label{eq_app_2m_inplane}
\end{eqnarray}
Starting from the two-magnon continuum
$S^{\zeta\zeta}_{2\cal{M}}(\bm{k},\omega)$ in Eq.\
(\ref{eq_app_2m_correlations}) calculated in the rotating
reference frame the wave vector values $\bm{k}$ are shifted by
$+\bm{Q}$ and $-\bm{Q}$. This gives two continua, their
superposition is the two-magnon scattering in the fixed reference
frame in Eq.\ (\ref{eq_app_2m_inplane}). Fig.\
\ref{fig_1and2m_kmq}(a) shows a plot of the first term in Eq.\
(\ref{eq_app_2m_inplane}) as a function of energy $\omega$ and
wave-vector $\bm{k}$ along the $\bm{b}^*$ direction. In the figure
the density of points indicates scattering intensity, which
decreases rapidly with increasing energy. The energy-integrated
two-magnon scattering intensity is plotted in Fig.\
\ref{fig_1and2m_kmq}(b)(solid points): the intensity is largest
near the magnetic Bragg peak position and cancels at the
ferromagnetic zone center $\bm{k} \rightarrow \bm{\tau}$. For
$S$=1/2 and $\Delta S$=0.125 the two-magnon spectral weight is
throughout most of the Brillouin zone comparable to the weight of
the in-plane polarized one-magnon mode $\omega^{-}_{\bm{k}}$
(dashed-dotted line) that occurs near the lower boundary of the
two-magnon continuum, whereas for wavevectors near the magnetic
Bragg peak position most scattering weight is in the one-magnon
channel.

Including the polarization factors for magnetic scattering the
one-magnon cross-section measured by neutron scattering is
\begin{equation}
I_{1\cal{M}}(\bm{k},\omega)= p_x
S^{xx}_{1\cal{M}}(\bm{k},\omega)+p_z
S^{zz}_{1\cal{M}}(\bm{k},\omega) \label{eq_app_1m}
\end{equation}
and the two-magnon cross-section is
\begin{equation}
I_{2\cal{M}}(\bm{k},\omega)=p_x S_{2\cal{M}}^{xx}({\bm k},\omega)
\label{eq_app_2m}
\end{equation}
where the polarization factors are
\begin{eqnarray}
p_x & = & 1+\cos^2\alpha_{\bm k} \nonumber \\
p_z & = & \sin^2\alpha_{\bm k}
\label{eq_app_polarizations}
\end{eqnarray}
The relative intensity of in-plane ($\parallel x$) and
out-of-plane ($\parallel z$) polarized excitations can therefore
be changed by varying the angle $\alpha_{\bm{k}}$ between the
scattering wave vector $\bm{k}$ and the axis $\hat{\bm z}$ normal
to the rotation plane. For Cs$_2$CuCl$_4$ the ordered spins rotate
nearly in the ($bc$) plane, so the axes are identified as
($x,y,z$) $\equiv$ ($b,c,a$). For comparison with experiments the
magnetic form factor $|f(\bm{k})|^2$ for Cu$^{2+}$ ions
\cite{Wilson95} was also included on the r.h.s. in Eqs.\
(\ref{eq_app_1m}) and (\ref{eq_app_2m}).

\appendix
\section{Square-lattice notation}
\label{app_sl}
Theoretical models\cite{Chung01,Zhou02} for the triangular lattice
in Fig.\ \ref{fig_cs2cucl4_structure}(b) often regard the problem
as a square-lattice with exchange $J^{\prime}$ is along the edges
and $J$ along one of the diagonals. A contour plot of the 2D
dispersion relation in Eq.\ (\ref{eq_dispersion}) using the
square-lattice basis\cite{sl} is shown in Fig.\ \ref{fig_sl}.



\begin{thebibliography}{10}
\bibitem{sl}
    The transformation of wavevector components from rlu units
    $\bm{k}=(h,k,l)$ to the square-lattice representation of
    Fig.\ \ref{fig_cs2cucl4_structure}(b) ($J^{\prime}$ along
    the square edges) is $k_x=(k-l)\pi$ and $k_y=(k+l)\pi$
    (for unit square-lattice spacing).
    The Fourier transfor of the exchange couplings is
    $J_{\bm{k}}=J\cos(k_x+k_y)+J^{\prime}(\cos k_x +\cos k_y)$,
    the incommensurate ordering wavevector is
    $\bm{Q}=(\frac{\pi}{2}+\epsilon_0 \pi ,\frac{\pi}{2}+ \epsilon_0 \pi)$
    and ``saddle" points of the 2D dispersion relation in Eq.\ (\ref{eq_dispersion})
    occur near
    $(k_x,k_y)=(\frac{3\pi}{4},\frac{3\pi}{4})+(-l,l) \pi$, with
    $l$ an even integer.
\end{thebibliography}
\end{document}